\documentclass[journal,10pt]{IEEEtran}

\usepackage{graphicx}
\usepackage{amssymb}
\usepackage{amsmath}

\usepackage{enumerate}

\usepackage{mathrsfs}
\usepackage[usenames, dvipsnames]{color}
\usepackage{float}
\usepackage{subfigure}
\usepackage{epstopdf}
\usepackage{mathrsfs}
\usepackage{caption}

\usepackage{lipsum}
\usepackage{algorithmic}

\newtheorem{theorem}{Theorem}

\newtheorem{proposition}{Proposition}
\newtheorem{lemma}{Lemma}

\newtheorem{example}{Example}
\newtheorem{definition}{Definition}

\newtheorem{remark}{Remark}

\newcommand{\ignore}[1]{}

\allowdisplaybreaks

\begin{document}
\renewcommand\thepage{}
\title{\LARGE \bf Coalitional Game Framework for Content Distribution Using Device-to-device Communication}

\author{Aditya MVS, Chitrarth Shrivastava and Gaurav S. Kasbekar}

\maketitle
{\renewcommand{\thefootnote}{} \footnotetext{ Aditya MVS is a post doctoral associate in University of Minnesota, Minneapolis, US, C. Shrivastava is with Goldman Sachs, Bengaluru, India and G. S. Kasbekar is with the Department of Electrical Engineering, Indian Institute of Technology Bombay, Mumbai, India. Their email addresses are vmurakon@umn.edu, chitrarth.shrivastava@ny.email.gs.com and gskasbekar@ee.iitb.ac.in. Aditya MVS and C. Shrivastava worked on this research while they were at IIT Bombay.}}

{\renewcommand{\thefootnote}{} \footnotetext{A preliminary version of this paper appeared in Proc. of IEEE VTC2019-Spring~\cite{RF:VTC:2019}.}}

\begin{abstract}
We consider a set of cellular users associated with a base station (BS) in a cellular network that employs Device-to-device (D2D) communication. A subset of the users request for some files from the BS.  Now, some of the users can potentially act as relays and forward the requested files, or partitions of files, from the BS to some of the requesting users (destination nodes) over D2D links. However, this requires cooperation among the cellular users. Also, when cellular users cooperate with each other, the total amount of energy consumed in transferring the requested files from the BS to the destination nodes can usually be considerably reduced compared to the case when each user separately downloads the file it needs from the BS. In this paper, we seek conditions under which users have an incentive to cooperate with each other. To this end, we model the above scenario using the frameworks of cooperative game theory and  stable partitions in coalitional games. We consider two different models for file transfer within a coalition: (i) Model A, in which the BS can split a file into multiple partitions and send these partitions to different relays, which multicast the partitions to the destination nodes of the coalition, and (ii) Model B, in which for each file, the BS sends the entire file to a single relay, which multicasts it to the destination nodes of the coalition. First, we explore the question of whether it is beneficial for \emph{all} the cellular users  to cooperate, i.e., whether the grand coalition is stable. For this we use the solution concept of \emph{core} from cooperative game theory. We 
show that, in general, the above coalitional game under Model A may have an empty core, i.e., it may not be possible to stabilize the grand coalition.  Next, we provide conditions under which 1) the core is always non-empty and 2) a \emph{$\mathbb{D}_c$-stable partition} always exists. Also, we show that under Model B, the problem of assigning relays to destination nodes so as to maximize the sum of utilities of all the users is NP-Complete. Finally, we show via numerical computations that a significant reduction in the energy expenditure of cellular users can be achieved via cooperation.
\end{abstract}

\section{Introduction}
\label{sec:introduction}
The demand for data in cellular networks has seen an explosive growth over the past decade.  As per the white paper released by CISCO~\cite{RF:Cisco}, the amount of global mobile data demand will increase seven-fold between 2016 and 2021. A straightforward solution is to increase the cell density in the congested areas of the network, thereby increasing network capacity~\cite{RF:Yeh}. However, this will also result in increased capital and operational costs to cellular operators. One alternative to avoid this is to use the concept of \emph{Device-to-device (D2D) communication} to improve the performance of the network~\cite{RF:D2D:survey}. When D2D communication is used, a base station (BS) can use some of its associated cellular users as \emph{relays} to forward data to other users over D2D links. Also, often multiple users request the same file (\emph{e.g.}, a popular file); in this case, a relay can \emph{multicast} the file, or a partition of the file, over D2D links to some of the nearby users requesting it \cite{RF:Z:Chang:2}. When such relaying is employed, the total amount of energy consumed in transferring the requested files from the BS to the requesting users (henceforth called destination nodes) can usually be considerably reduced compared to the case when each destination node separately downloads the file it needs from the BS~\cite{RF:Andreev},~\cite{RF:Andreev1}. Such offloading of data by the BS to D2D links can also improve the capacity of the cellular network~\cite{RF:Andreev},~\cite{RF:Andreev1}.

We consider a set of cellular users associated with a BS in a cellular network that employs D2D communication. A subset of the users request for some files from the BS.  As mentioned above, some of the users can potentially act as relays and forward the requested files, or partitions of files, from the BS to the destination nodes over D2D links, which results in greater energy efficiency. However, this requires \emph{cooperation} among the cellular users.     Since the energy efficiency of D2D communication decreases with the increase in distance between the communicating users~\cite{RF:Energy,RF:Energy:2}, two cellular users who are located far away from each other may be better off downloading the content they need directly from the BS instead of cooperating with each other. For example, consider five cellular users $\{1,2,3,4,5\}$ requesting the same file from the BS. Users $1,2$ are located close to each other, users $3,4,5$ are located close to each other and the subsets of users $\{1,2\}$ and $\{3,4,5\}$ are located far away from each other. In this case, users $1$ and $2$ (respectively, users $3,4$ and $5$) have an incentive to cooperate among themselves, e.g., user $1$ may download the file from the BS and send it over a D2D link to user $2$ (respectively, user $4$ may download the file from the BS and multicast it over D2D links to users $3$ and $5$).

A set of cellular users who cooperate among themselves is called a \emph{coalition} \cite{RF:Osborne}. As the above example shows, a set of cellular users would  cooperate with each other only when they benefit from this cooperation. In this paper, we investigate conditions under which users have incentives to cooperate among themselves to form a coalition; we also study the problem of assigning relays to destination nodes  so as to minimize the total energy consumed in transferring files from the BS to destination nodes via relays within a coalition. We model the above problem using the framework of \emph{cooperative game theory}~\cite{RF:Osborne} and \emph{stable partitions} in coalitional games~\cite{RF:stable:coalitions}.
We consider two different models for relaying within a given coalition: (i) Model A, in which the BS can split a file into multiple partitions and send these partitions to different relays, which multicast the partitions to the destination nodes, and (ii) Model B, in which for each file, the BS sends the entire file to a single relay, which multicasts it to the destination nodes.   The results for Model A are divided into two parts:  
 \begin{itemize}
 \item
In the first part, we investigate conditions under which it is beneficial for \emph{all} the cellular users  to cooperate, i.e., the grand coalition is stable. For this we use the solution concept of \emph{core} \cite{RF:Osborne} from cooperative game theory. We show that, in general, the above coalitional game under Model A may have an empty core, \emph{i.e.}, it may not be possible to stabilize the grand coalition.  Next, we consider  an important special case of this game, wherein all D2D and BS-cellular user communication links are symmetric across cellular users and the D2D data rates are much higher than the BS-cellular user data rates. Such a scenario would occur in practice when  all the cellular users are located close to each other, \emph{e.g.}, in a stadium or concert hall, and hence data exchange between a pair of users can occur at a fixed and high rate, but the users are situated far away from the BS. In this special case, we show that the core is always non-empty. 
\item
\ignore{
Now, the well-known \emph{merge and split algorithm} \cite{RF:stable:coalitions} can be used to find a partition of a set of users into multiple coalitions.  It is known that in general, the merge and split algorithm  converges to an arbitrary partition, and only under special conditions (when a strictly \emph{$\mathbb{D}_c$-stable partition} exists), it converges to the partition with minimum total expended energy \cite{RF:stable:coalitions}.}

In the second part, we study the case where the set of cellular users can be partitioned into groups, with each group of users forming a coalition to cooperate among themselves. In particular, we present a set of sufficient conditions under which a strictly \emph{$\mathbb{D}_c$-stable partition}~\cite{RF:stable:coalitions} exists. From the results shown in \cite{RF:stable:coalitions}, it can be concluded that the $\mathbb{D}_c$-stable partition corresponds to the partition with minimum total expended energy by the cellular users in transferring files from the BS to destination nodes via relays; also, a simple \emph{merge and split algorithm}~\cite{RF:stable:coalitions} converges to this partition. A special case where this happens is when the  cellular users are located in the form of clusters, such that cellular users of the same cluster are located close to each other and cellular users of different clusters are located far away from each other. 
\end{itemize}
Next, we show that under Model B, the problem of assigning relays to destination nodes so as to maximize the sum of utilities~\footnote{The utility of a user is a function of the valuation it derives from the file it downloads, if any, and the cost it incurs due to the energy consumption during communication.}  of all the users is NP-Complete. Hence, we do not perform a cooperative game theoretic analysis of Model B. Instead, we provide heuristics to solve the utility maximization problem in this model and evaluate their performance via numerical computations.

The rest of the paper is organised as follows. We present a review of related prior literature in Section~\ref{sec:related:work:coalition}. In Section~\ref{SC:system:model}, we describe our network model.  We investigate conditions under which the core is non-empty under Model A in Section~\ref{SC:model:A}. In Section~\ref{SC:coalition:formation}, we present  conditions under which a $\mathbb{D}_c$-stable partition exists under Model A. In Section~\ref{SC:model:B}, we prove that under Model B, the problem of assigning relays to destination nodes so as to maximize the sum of utilities of all the users is NP-complete  and provide heuristics for assignment of relays to destination nodes. In Section~\ref{SC:numerical:results}, we show via numerical simulations that the total energy expended by cellular users is significantly reduced through cooperation. Finally we conclude our work in Section~\ref{SC:conclusions:coalition}. 

\section{Related Work}
\label{sec:related:work:coalition}
We now review related prior literature. Resource allocation in cooperative cellular networks with the objective of achieving energy efficiency is a well-studied problem. A cooperative cellular network wherein the BS sends content to some selected cellular users, which in turn multicast it to other cellular users is considered in~\cite{RF:Yuan}. The problem of joint optimization of the cost of the energy consumption and cellular-link usage in the network by appropriately selecting the transmission rates of the relays is studied.  In \cite{RF:Z:Chang} \cite{RF:Z:Chang1}, cellular data offloading in a cooperative cellular network, in which data transmission to the end users and energy harvesting are simultaneously performed, is studied. Algorithms to optimally schedule the data offloading and radio resources in order to maximize the energy efficiency of the network are presented. A cooperative framework in a cellular network where the BS transmits a file only once to a cellular user, which in turn relays it over D2D links  to all the other users that request it  is studied in~\cite{RF:Yaacoub}. In~\cite{RF:Y:Zhang}, the content distribution problem in a cooperative network, wherein the BS selects relays to broadcast some content, is modeled  as a non-transferable utility coalition formation game, in which the utility function takes into account energy efficiency and mutual interference among multiple relays. A distributed algorithm is presented using which cellular users can self-organize among themselves to form coalitions. A coalition formation game is also studied in~\cite{RF:L:Kanj,RF:coalition:1}. In the model in~\cite{RF:L:Kanj}, cellular users can cooperate and self-organize to form coalitions among themselves and use them to distribute content. In each coalition, a cellular user acts as the head of the coalition; it receives data from the BS and then  multicasts the data to the users in the coalition. A distributed algorithm for coalition formation is proposed and the energy efficiency when coalitions are formed using this algorithm is shown to be higher than that in a non-cooperative cellular network and in a cellular network where all the cellular users requesting the content form a grand coalition. A centralized coalition formation game is formulated as a mixed integer linear program (MILP) in~\cite{RF:coalition:1}. Since the formulated MILP is NP-hard, a linear approximation of the above problem is solved to find coalitions among the cellular users. The distribution of location specific content among users, which preserves privacy by securing the location information is studied in~\cite{RF:privacy}. The human factor of willingness to share content is considered in~\cite{RF:Pan}. Cellular users (represented by humans) are grouped in terms of the type of content they request, and each group has different willingness factors of sharing, which are represented by different probabilities of sharing.  

In this paper, we use the framework of coalition formation proposed in~\cite{RF:stable:coalitions}. This framework has been widely adapted for several wireless communication and social network related problems such as interference management in visible light communication networks~\cite{RF:Chen},  community identification in dynamic social networks~\cite{RF:Xiao}, cooperation of single antenna devices to form virtual multiple antenna systems~\cite{RF:Saad:1}, and in collaborative spectrum sensing~\cite{RF:Saad:2}. Coalitions are formed using the merge and split algorithm. The partitioning of users using the merge and split algorithm in a cloud based radio access network where baseband units are in a centralised location and radio heads are distributed across multiple sites is studied in~\cite{RF:Taleb}. Application of the merge and split algorithm to a coalition formation game for content distribution in cellular networks is studied in~\cite{RF:coalition:1},~\cite{RF:coalition:3}. However, in \cite{RF:coalition:3}, the D2D communication occurs underlay and hence an efficient policy to allocate resources is studied taking interference into consideration; in contrast, in our model, D2D communication occurs overlay.   

Also, none of the above papers study conditions under which the grand coalition is stable; nor do they study conditions under which a strictly $\mathbb{D}_c$-stable partition exists. To the best of our knowledge, \emph{our work is the first to use a coalitional game framework to study conditions under which the grand coalition is stable and those under which a $\mathbb{D}_c$-stable partition exists} in the context of content distribution in a cellular network employing D2D communication. Our analysis provides insight into conditions under which it is, and is not, beneficial for all the cellular users to cooperate. Also, the significance of the conditions that we have derived for a $\mathbb{D}_c$-stable partition to exist is that if these conditions are met, then the merge and split algorithm can be used to find the partition with minimum total
expended energy. In addition, although the merge and split algorithm has an exponential running time in general, \emph{if the conditions presented in this paper are satisfied, then the merge and split algorithm proposed in~\cite{RF:coalition:3} converges to the partition with minimum total expended energy in polynomial time.} 

\section{Network Model}
\label{SC:system:model}
We consider a single cell containing a base station (BS) and multiple cellular users. Let $\mathcal{N}=\{1,\ldots,N\}$ denote the set of all cellular users. In a given time slot, some of them  request for some files from the BS-- we refer to such cellular users as ``destination nodes''. We assume that each destination node requests for exactly one file.  Let $\mathcal{M}=\{1,\ldots,M\}$ denote the set of all requested files. The BS seeks to reduce the energy consumption by employing some of the cellular users as relays to forward the requested files to the destination nodes over D2D links, instead of directly sending the requested file to each destination node. (A destination node of a file may also act as a relay for the same and/ or other files). To this end, the BS divides the set $\mathcal{N}$ into multiple groups of cellular users (\emph{e.g.}, a group may be a set of cellular users located close to each other)  such that the cellular users within each group cooperate among themselves to download their requested files. Each such group is called a \emph{coalition}.
\begin{definition}
\label{DN:coalition}
A \emph{coalition} $S \subseteq \mathcal{N}$ is a set of users who cooperate among themselves. We refer to $\mathcal{N}$ as the \emph{grand coalition}~\cite{RF:Osborne}.
\end{definition}

In each coalition $S$, the BS employs some of the cellular users as relays to forward the requested files to the destination nodes in $S$ over D2D links. If a file (\emph{e.g.}, a popular file) is requested by more than one destination node in a coalition $S$, then each relay in $S$ \emph{multicasts} the file, or a partition of the file it received from the BS, to \emph{all} the destination nodes in $S$ that requested the file. Based on how a file is distributed within a coalition, we consider two different models:
\begin{enumerate}[(A)]
\item 
the BS can split a file into multiple partitions and send these partitions to different relays in a coalition, which multicast the partitions to the destination nodes of that coalition,
\item
for each file, the BS sends the entire file to a single relay in a coalition, which multicasts it to the destination nodes of that coalition.
\end{enumerate}
In the sequel, we refer to the above models as Model A and Model B; Model A is studied in Sections~\ref{SC:model:A} and~\ref{SC:coalition:formation} and Model B in Section~\ref{SC:model:B}.
Let $\alpha_{i,m}$, $i\in \mathcal{N},m\in \mathcal{M}$ denote the fraction of file $m$ that is sent by the BS to relay $i$. In Model A (respectively, Model B), these variables must satisfy $\alpha_{i,m} \in [0,1] \ \forall i\in \mathcal{N},m\in \mathcal{M}$ (respectively, $\alpha_{i,m} \in \{0,1\} \ \forall i\in \mathcal{N},m\in \mathcal{M}$). Also, in both models, the following must be satisfied for every coalition $S$:
\begin{equation}
\sum_{i \in S} \alpha_{i,m} =
\begin{cases}
1\hspace{5mm} \mbox{if a user in $S$ requests file $m$},\\
0\hspace{5mm} \mbox{otherwise.}
\end{cases}
 \end{equation}

We assume that the BS knows the channel conditions between itself and different cellular users and among different cellular users through Channel State Information (CSI) conveyed by the cellular users. This information can be estimated using \emph{reference signals}, which are sent at known transmit powers and whose received powers are measured at the receivers~\cite{RF:channel:estimation}. Using the channel conditions, the data rates that can be achieved between different pairs of devices can be found. Also, we focus on low mobility scenarios, in which channel conditions change slowly with time; hence, to a good approximation, it can be assumed that the channel conditions remain the same throughout the duration of a time slot. A similar assumption has been made in many prior works \cite{RF:Yuan, RF:Z:Chang, RF:Z:Chang1, RF:Yaacoub}.
Let $R_{s,i}$ denote the achievable data rate between the BS and relay $i$ and $P_{Rx,i}(R_{s,i})$ denote the power used by relay $i$ to receive a file from the BS.   We formulate the energy spent by users in the cellular network using a model similar to that in~\cite{RF:L:Kanj}. Consider a relay $i$ in coalition $S$. The total energy spent by relay $i$ in receiving (partitions of) files from the BS is given by:
\begin{equation}
\label{EQ:E:r:j}
E_{s,i}(S)=\sum\limits_{m=1}^M\frac{\alpha_{i,m}X_m}{R_{s,i}}P_{s,i}(R_{s,i}),
\end{equation}
where $X_m$ denotes the size of file $m \in \mathcal{M}$. Equation \eqref{EQ:E:r:j} holds because $\alpha_{i,m}X_m$ is the number of bits of file $m$ downloaded by relay $i$ from the BS, $\frac{\alpha_{i,m}X_m}{R_{s,i}}$ is the amount of time spent and $\frac{\alpha_{i,m}X_m}{R_{s,i}}P_{s,i}(R_{s,i})$ is the energy consumed during the download of bits of file $m$ to relay $i$ from the BS. Let $R_{D2D,i,j}$ denote the data rate at which relay $i$ can transmit (a partition of) a file to the destination node $j$ over a D2D link and let $P_{Tx,i,j} (R_{D2D,i,j})$ be the transmission power $i$ uses. If multiple destination nodes in $S$ request a file, say $m$, then a relay, say $i$, multicasts (a partition of) the file to these destination nodes at the rate $R_{D2D,i,S_m}=\min_{j\in S_m,j\neq i}\{R_{D2D,i,j}\}$\footnote{When $S_m\setminus\{i\}$ is an empty set, we define the value of the $\min$ function to be any arbitrary positive number.}, where $S_m\subseteq S$ is the set of destination nodes in $S$ which request file $m$. Similar to \eqref{EQ:E:r:j}, the total energy spent by relay $i$ in multicasting (partitions of) files in $\mathcal{M}$ to destination nodes is:
\begin{equation}
\label{EQ:E:t:j}
E_{t,i}(S)=\sum\limits_{m=1}^M\frac{\alpha_{i,m}X_m}{R_{D2D,i,S_m}}P_{Tx,i,S_m}(R_{D2D,i,S_m})d_m(S_m\setminus\{i\}),
\end{equation}
where 
\begin{equation*}
d_m(A)=
\begin{cases}
1,\hspace{2mm}\mbox{if at least one user in $A$ requests file } m,\\
0,\hspace{2mm}\mbox{otherwise.}
\end{cases}
\end{equation*}
and $P_{Tx,i,S_m}(R_{D2D,i,S_m})$ is the power required by relay $i$ to multicast file $m$ to the destination nodes in $S_m$ over D2D links at the rate $R_{D2D,i,S_m}$.  Now, for every user $i\in \mathcal{N}$ which acts as a relay, energy is consumed in receiving data from the BS and multicasting it to destination nodes. However, since only a limited amount of battery energy is available with a relay, we let $E_i$ denote the maximum amount of energy that may be spent by user $i$ on relaying activities in the given time slot. Each relay $i\in S$ must satisfy the following constraint:
\begin{equation}
\label{EQ:relay:max:energy:constraint}
E_{s,i}(S)+E_{t,i}(S) \leq E_i.
\end{equation}
\ignore{
Recall that a relay multicasts a file, or a partition of a file, to all the destination nodes that request the file. In particular, let $\mathcal{N}_i$ denote the set of destination nodes which request file $i$. A relay $j$ which multicasts a fraction $\alpha_{i,j}$ of file $i$ multicasts it at data rate $R_{Tx,j}^i=\min\{R_{j,l}:l\in \mathcal{N}_i\}$,  \emph{i.e.}, at the minimum of the data rates from relay $j$ to each of the destination nodes that request file $i$. }

Also, each destination node which requests a file must spend some energy on receiving partitions of the file from different relays. If a destination node $i\in S$ requests file $m$, then the energy it spends in receiving (partitions of) file $m$ is given by:
\begin{equation}
\label{EQ:E:m:j}
E_{r,i}^m(S)=\sum\limits_{j\neq i}\frac{\alpha_{j,m}X_m}{R_{D2D,j,S_m}}P_{Rx,j,i}(R_{D2D,j,S_m}),
\end{equation} 
where $P_{Rx,j,i}(R_{D2D,j,S_m})$ is the power required by destination node $i$ while receiving data from relay $j$ at rate $R_{D2D,j,S_m}$.

Let $C_{s,i}(S)$ (respectively, $C_{t,i}(S)$) denote the monetary cost corresponding to the energy consumed at relay $i$, when it is in coalition $S$, due to the energy expenditure $E_{s,i}(S)$ (respectively, $E_{t,i}(S)$)  incurred by relay $i$ while downloading partitions of files from the BS (respectively, multicasting partitions of files to destination nodes). Similarly, let $C_{r,i}^m(S)$ be the monetary cost corresponding to the energy, $E_{r,i}^m(S)$, spent by destination node $i$ in receiving file $m$ that it requested from relays when it is a part of coalition $S$. We assume that these monetary costs are linear functions of the energy consumed: $C_{s,i} (S)=a E_{s,i}(S)$, $C_{t,i}(S)=a E_{t,i}(S)$ and $C_{r,i}^m(S)=a E_{r,i}^m(S)$, where $a$ is a  constant. Next, let:
\begin{equation*}
d_{i,m}=
\begin{cases}
1\hspace{2mm}\mbox{if the user $i$ requests file $m$},\\
0 \hspace{2mm}\mbox{otherwise}.
\end{cases}
\end{equation*}
Note that $d_{i,m}$, $i \in \mathcal{N}$, $m \in \mathcal{M}$, are \emph{constants} that are known a priori. Let $C_i(S) = C_{s,i} (S) + C_{t,i}(S) + \sum\limits_{m=1}^Md_{i,m}C_{r,i}^m(S) = a(E_{s,i} (S) + E_{t,i}(S) + \sum\limits_{m=1}^Md_{i,m}E_{r,i}^m(S))$.  If destination node $i$ requests file $m$, we let $U_{i,m}$  denote the valuation that destination node $i$ derives from file $m$. The utility of user $i$ is defined to be the difference between the valuation that it derives from the file that it requests and the costs due to the energy consumption during communication, \emph{i.e.}:  
$\sum\limits_{m=1}^M d_{i,m}(U_{i,m}-C_{r,i}^m(S))-C_{s,i}(S)-C_{t,i}(S)$. Also, for a coalition $S$ and file $m \in \mathcal{M}$, let $E^m(S)$ be the total energy consumed in transferring file $m$ from the BS to the destination nodes in $S$ that request for file $m$. Then:
\begin{align}
\label{eq:cost:file}
E^m(S)&=\sum\limits_{i\in S}\alpha_{i,m}X_m\Bigg(\Bigg(\frac{P_{s,i}(R_{s,i})}{R_{s,i}}+\frac{P_{Tx,i,S_m}(R_{D2D,i,S_m})}{R_{D2D,i,S_m}}\Bigg)\nonumber\\&\quad+\sum\limits_{j\neq i}\frac{\alpha_{j,m}X_m}{R_{D2D,j,S_m}}P_{Rx,j,i}(R_{D2D,j,S_m})d_{i,m}\Bigg)
\end{align} 

Hence, when a group of cellular users form a coalition $S$, the following optimization problem maximizes their sum of utilities:
\begin{align}
\label{EQ:PS}
P(S): \max\limits_{\alpha_{i,m}}\sum\limits_{i\in S}\Bigg(\sum\limits_{m=1}^M d_{i,m}U_{i,m}-C_i(S)\Bigg)
\end{align} 
subject to:
\noindent\\
1) $\alpha_{i,m}\geq 0, \ \forall i\in \mathcal{S}, m\in \mathcal{M}$,\\
2) $\sum\limits_{i\in S}\alpha_{i,m}= d_m(S), \ \forall m\in \mathcal{M}$,\\
3) $E_{s,i}(S)+E_{t,i}(S)\leq E_i, \ \forall i\in S$.\\

Constraint 1) says that the variables $\alpha_{i,m}$ must be non-negative, constraint 2) says that if a user in coalition $S$ requests file $m$, then the entire file must be downloaded from the BS by the relays in coalition $S$ and constraint 3) says that the amount of energy consumed by each user $i\in S$ due to its relaying services must not exceed $E_i$. In addition, in Model A (respectively, Model B), the constraint $\alpha_{i,m} \leq 1$ (respectively, $\alpha_{i,m} \in \{0,1\}$) must be met for all $i \in S$ and $m \in \mathcal{M}$. 

Note that in the above network model, the total amount of energy required to transfer files from the BS to all the requesting destination nodes can usually be considerably reduced when cellular users \emph{cooperate} with each other, transfer files by relaying and transfer payments among themselves (\emph{e.g.}, payments may be transferred from a destination node to the relays that forward data to it), as compared to the case when each destination node separately downloads the file it needs from the BS. Hence, we are interested in finding conditions under which it is beneficial for the cellular users of the network to cooperate with each other. Specifically, we consider two cases in this work: 1) when the set of \emph{all} cellular users in $\mathcal{N}$ cooperate among themselves (see Sections~\ref{SC:model:A} and~\ref{SC:model:B}), and 2) when the set of cellular users is partitioned into groups (coalitions) such that the users of each coalition cooperate among themselves (see Section~\ref{SC:coalition:formation}).

\section{Cooperative Game Theoretic Analysis of Model A}
\label{SC:model:A}
In this section, we perform a cooperative game theoretic analysis of Model A, in which the variables  $\alpha_{i,m}$, $i \in \mathcal{N}, m \in \mathcal{M}$ may take real values in $[0,1]$ and $\sum\limits_{i \in S} \alpha_{i,m}=d_m(S)$ for all $m\in \mathcal{M}$.

\subsection{Mathematical Preliminaries}
We first define some terminology and notations of cooperative game theory, which we use in the sequel. 
\begin{definition}
A \emph{coalitional game with transferable payoffs} $(\mathcal{N},v)$ consists of a set, $\mathcal{N}$,  of $N$ users and a real-valued function $v(\cdot)$ associated with each coalition $S\subseteq \mathcal{N}$.  $v(S)$ is called the \emph{value} of the coalition $S$~\cite{RF:Osborne}.
\end{definition}

In our work, we define the optimal (maximum) value of the objective function in \eqref{EQ:PS} to be the value, $v(S)$, of the coalition $S$. In this section, we are particularly interested in conditions under which it is beneficial for \emph{all} the cellular users in $\mathcal{N}$ to cooperate, \emph{i.e.}, the grand coalition (see Definition~\ref{DN:coalition}) is stable. For this we use the solution concept of \emph{core} from cooperative game theory~\cite{RF:Osborne}. 
\begin{definition}
Let $(\mathcal{N},v)$ be a coalitional game with transferable payoffs. A vector $(x_j)_{j\in \mathcal{N}}$ is said to be a \emph{feasible payoff profile} if $x(\mathcal{N})=\sum\limits_{j=1}^Nx_j=v(\mathcal{N})$. The \emph{core} is the set of all feasible payoff profiles $(x_j)_{j\in\mathcal{N}}$ for which $x(S)=\sum\limits_{j\in S}x_j\geq v(S)$ for every coalition $S \subseteq \mathcal{N}$~\cite{RF:Osborne}.
\end{definition}

Now, \emph{the grand coalition can be stabilized iff the core is non-empty}~\cite{RF:Osborne}. This can be explained as follows. Recall that the  value, $v(\mathcal{N})$, of the grand coalition is given by the maximum value of the objective function in  \eqref{EQ:PS} with $S = \mathcal{N}$.   Suppose the core is non-empty and this value $v(\mathcal{N})$ is shared among the users in $\mathcal{N}$ as per an element $x$ in the core, \emph{i.e.}, payments are transferred among the users in $\mathcal{N}$ such that the overall utility of user $j \in \mathcal{N}$ (taking into account payments made by and to $j$) becomes $x_j$. Note that this can be done since $x(\mathcal{N})=\sum\limits_{j=1}^Nx_j=v(\mathcal{N})$. Then no subset of users $S \subseteq \mathcal{N}$ has an incentive to split from the grand coalition, \emph{i.e.}, the grand coalition is stable. To prove this, suppose a subset of users $S \subseteq \mathcal{N}$ formed a separate coalition and shared their value, $v(S)$, as per the vector $y$. However, it would be in the interest of user $j \in S$ to split from the grand coalition only if $y_j > x_j$. Hence, $v(S) = \sum_{j \in S} y_j > \sum_{j \in S} x_j = x(S)$, which contradicts the fact that $x$ is an element of the core. Conversely, it is easy to see that if the core is empty, then there would always be an incentive for some subset $S \subseteq \mathcal{N}$ to split from the grand coalition, regardless of how the value $v(\mathcal{N})$ is shared among the users of $\mathcal{N}$; \emph{i.e.}, the grand coalition cannot be stabilized.  

So, the grand coalition can be stabilized iff the core is non-empty. Hence, in this section, we seek conditions under which the core is non-empty. First, in Section~\ref{SSC:model:A:general}, we show that, \emph{in general, the above coalitional game under Model A may have an empty core}. Next, in Section~\ref{SSC:model:A:special}, we show that \emph{in an important special case of this game, the core is always non-empty.} 

\subsection{General Game}     
\label{SSC:model:A:general}
Consider the above coalitional game under Model A. First, note that since the variables $\alpha_{i,m}$, $i \in \mathcal{N}, m \in \mathcal{M}$ may take real values, the optimization problem $P(S)$ defined in Section~\ref{SC:system:model} (see \eqref{EQ:PS}) is a \emph{linear program} and hence can be optimally solved in polynomial time~\cite{RF:karmarkar:LP:poly:time}. Hence, the value, $v(S)$, of each coalition $S$, which is the optimal value of the objective function in \eqref{EQ:PS}, can be found in polynomial time.

The following example shows that this coalitional game may have an empty core.
\begin{example}
\label{EG:empty:core}
Suppose $\mathcal{N} = \{1,2,3,4,5,6\}$. Consider two coalitions $S_1=\{1,2,3\}$ and $S_2=\{4,5,6\}$. Note that $S_1\cup S_2= \mathcal{N}$ and $S_1\cap S_2=\emptyset$. Suppose users $1,2,4,5$ request file $1$, which has a size of $X_1 = 1$ and users $3,6$ request none. Also, suppose $R_{s,i}=1$ for $i\in\{1,2,4,5\}$, $R_{s,i}=8$ for $i\in\{3,6\}$, $P_{s,i}(\cdot)=P_{Tx,i,j}(\cdot)=P_{Rx,i,j}(\cdot)=1$ for all $i,j\in \mathcal{N}$. Suppose $R_{D2D,3,1}=R_{D2D,3,2}=R_{D2D,6,4}=R_{D2D,6,5}=8$, $R_{D2D,3,4}=R_{D2D,3,5}=R_{D2D,6,1}=R_{D2D,6,2}=1$ and $R_{D2D,1,2} =R_{D2D,1,4}=R_{D2D,1,5}=R_{D2D,2,4}=R_{D2D,2,5}=R_{D2D,4,5}=1$. Let $a = 1$ so that $C_i(S)=E_{s,i}(S)+E_{t,i}(S)+E_{r,i}^1(S)$ $\forall i\in\mathcal{N}$. 

In this example, we have:
\begin{align*}
v(S_1)&=U_{1,1}-C_1(S_1)+U_{2,1}-C_2(S_1)-C_3(S_1),\\
v(S_2)&=U_{4,1}-C_4(S_2)+U_{5,1}-C_5(S_2)-C_6(S_2),\\
v(\mathcal{N}) &=U_{1,1}-C_1(\mathcal{N})+U_{2,1}-C_2(\mathcal{N})-C_3(\mathcal{N})\\
&+U_{4,1}-C_4(\mathcal{N})+U_{5,1}-C_5(\mathcal{N})-C_6(\mathcal{N}).
\end{align*}
We will show that $v(S_1)+v(S_2) > v(\mathcal{N})$. 
Note that the sums of the $U_{i,1}$ components are equal in $v(S_1)+v(S_2)$ and $v(\mathcal{N})$ (equal to $U_{1,1} + U_{2,1} + U_{4,1} + U_{5,1}$). So next we consider the energy cost components. Consider coalition $S_1$ and let $i(S_1)$ denote the total energy consumed in distributing file 1 among users in $S_1$ when user $i\in S_1$ downloads the entire file from the BS. We now calculate the value of the term $1(S_1)$; user 1 consumes $1$ unit of energy (see \eqref{EQ:E:r:j}) to download the file from the BS, consumes $1$ unit of energy (see \eqref{EQ:E:t:j}) to transmit the file to user 2 and user 2 consumes 1 unit of energy (see \eqref{EQ:E:m:j}) in receiving the file from user 1. So the total energy cost $1(S_1)=3$. $2(S_1)$ and $3(S_1)$ can be calculated similarly and it can be seen that $2(S_1)=3$ and $3(S_1)=\frac{1}{2}$. Hence, it is easy to check that $C_1(S_1) + C_2(S_1) + C_3(S_1)$ is minimized when the BS  sends file $1$ to user $3$ and user $3$ multicasts it to users $1$ and $2$. Similarly, $C_4(S_2) + C_5(S_2) + C_6(S_2)$ is minimized when the BS sends file $1$ to user $6$ and user $6$ multicasts it to users $4$ and $5$. The minimum costs for $S_1$ (respectively, $S_2$) are as follows: $C_1(S_1)=C_2(S_1)=\frac{1}{8}$ and $C_3(S_1)=\frac{1}{4}$ (respectively, $C_4(S_2)=C_5(S_2)=\frac{1}{8}$ and $C_6(S_2)=\frac{1}{4}$). Thus the total cost term in $v(S_1)+v(S_2)$ is $C_1(S_1) + C_2(S_1) + C_3(S_1) + C_4(S_2) + C_5(S_2) + C_6(S_2) = 1$.

On the other hand, it can be shown that the total energy cost for the coalition $\mathcal{N} = S_1\cup S_2$ is minimized when the BS sends a fraction $\beta \in [0,1]$ of  file $1$ to user $3$ and a fraction $1-\beta$ to user $6$, and users $3$ and $6$ in turn  multicast the partitions they receive to  users $1,2, 4$ and $5$. The resultant cost terms are as follows: $C_1(\mathcal{N})=C_2(\mathcal{N})=C_4(\mathcal{N})=C_5(\mathcal{N})=1$ and $C_3(\mathcal{N})+C_6(\mathcal{N})=\frac{9}{8}$. Hence, the total cost term in the coalition $\mathcal{N}$ is $C_1(\mathcal{N})+C_2(\mathcal{N})+C_3(\mathcal{N})+C_4(\mathcal{N})+
C_5(\mathcal{N})+C_6(\mathcal{N}) = \frac{41}{8}$. Thus, the total cost term in the coalition $\mathcal{N}$, which is $\frac{41}{8}$,  is greater than the total cost term in  $v(S_1)+v(S_2)$, which is $1$. Hence, $v(S_1)+v(S_2) > v(\mathcal{N})$.  

Now, let $(x_j)_{j\in \mathcal{N}}$ be a feasible payoff profile in the core. Then we have $x(\mathcal{N})=v(\mathcal{N})$ and $x(S)\geq v(S)$ for every coalition $S$. Since $S_1\cap S_2=\emptyset$ and $S_1\cup S_2=\mathcal{N}$, we can write:
$v(\mathcal{N})=x(\mathcal{N}) = \sum_{j \in \mathcal{N}} x_j = \sum_{j \in S_1} x_j + \sum_{j \in S_2} x_j  = x(S_1)+x(S_2)
\geq v(S_1)+v(S_2)>v(\mathcal{N})$,
which is a contradiction. This proves that the core is empty. 
$\blacksquare$
\end{example}

Intuitively, the core is empty in the above example due to the following reason. The set of all cellular users, $\mathcal{N}$, consists of two disjoint clusters of users, $S_1$ and $S_2$, such that the achievable data rate between user $j_1$ and user $j_2$ is low  for every pair $j_1 \in S_1$ and $j_2 \in S_2$ (in particular, $R_{D2D,3,4}=R_{D2D,3,5}=R_{D2D,6,1}=R_{D2D,6,2}=1$). On the other hand, the achievable data rates among users within each cluster are high; in particular, $R_{D2D,3,1}=R_{D2D,3,2}=R_{D2D,6,4}=R_{D2D,6,5}=8$.  Recall that a relay multicasts data at the minimum achievable data rate between itself and any destination node within its coalition requesting the file.  So when coalition $S_1$ (respectively, $S_2$) separates from the other users in $\mathcal{N}$, data can be multicast at a high rate of $8$ from relay $3$ to users $1$ and $2$ (respectively, from relay $6$ to users $4$ and $5$), due to which the energy consumption is low. However, when all users in the grand coalition $\mathcal{N}$ cooperate, data has to be multicast at a low rate of $1$, due to which the energy consumption is high. Hence, the grand coalition cannot be stabilized in this example, \emph{i.e.}, it is not beneficial for all the users in $\mathcal{N}$ to cooperate.    

\ignore{
\begin{remark}
\label{RM:cohesive:game}
A coalitional game with transferable payoffs $(\mathcal{N},v)$ is \emph{cohesive} if: $v(\mathcal{N})\geq \sum\limits_{i=1}^k v(S_i)$, $\forall S_1, \ldots, S_k$ such that $\cup_{i=1}^k S_i=\mathcal{N}$ and $S_i\cap S_j=\phi$ for all $i\neq j$~\cite{RF:Osborne}. This means that the total value available to players to distribute among themselves is maximized when they \emph{all} cooperate. However, Example~\ref{EG:empty:core} shows that in general, the above coalitional game $(\mathcal{N},v)$ need not be a cohesive game. 
\end{remark}}

\subsection{Special Case}
\label{SSC:model:A:special}
\begin{figure}[!hbt]
\centering
\includegraphics[scale=0.3]{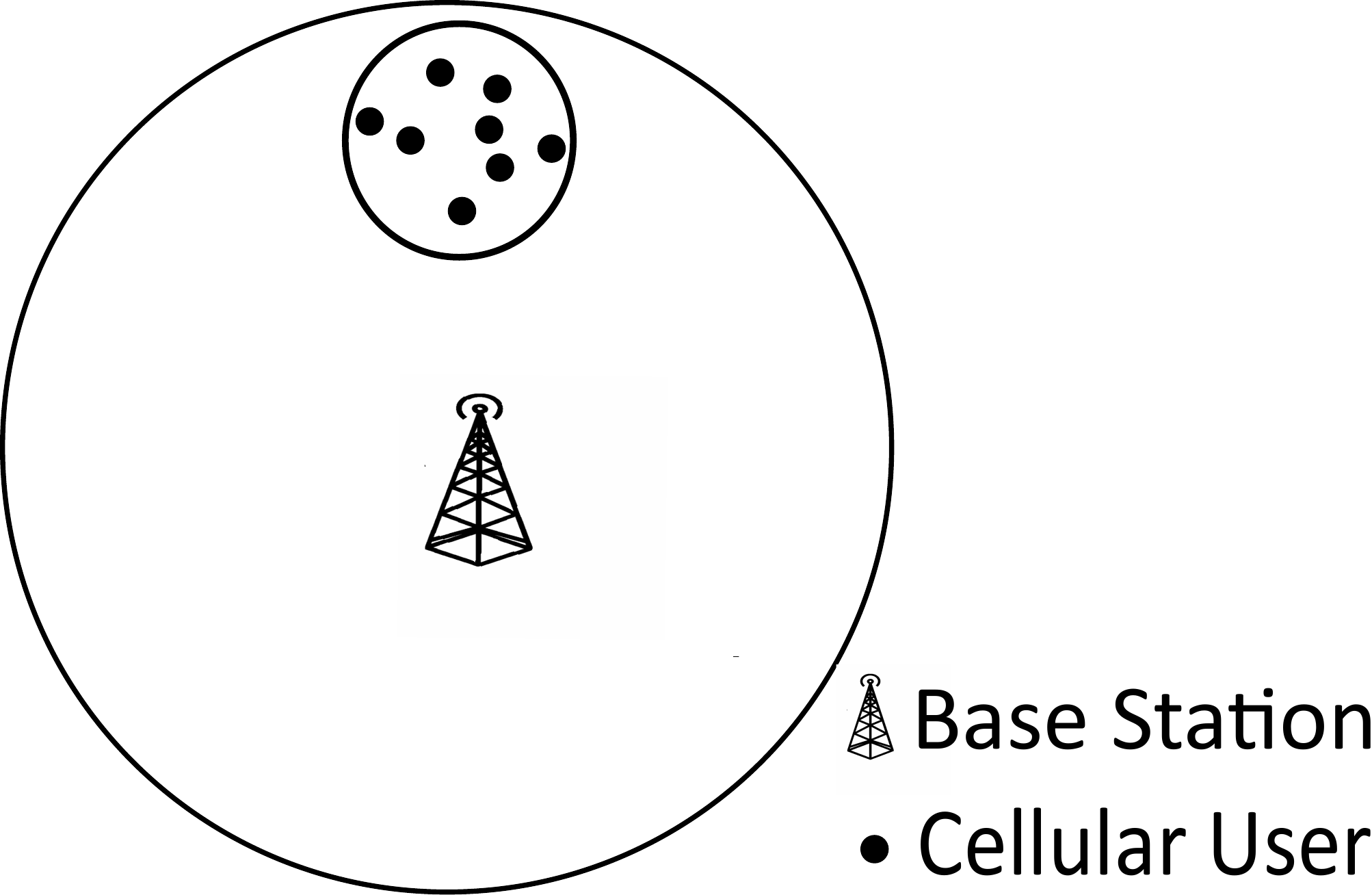}
\caption{The figure shows a cell with a BS located at the center and cellular users located close to each other, but far from the BS.}
\label{figure:core}
\end{figure}

We now analyse a special case of the coalitional game under Model $A$, in which all D2D and BS-cellular user links are symmetric across cellular users. Specifically, we assume that: (i) all D2D communications occur at a constant rate, say $R_{D2D}$, \emph{i.e.}, $R_{D2D,i,j} = R_{D2D}$ $\forall i, j \in \mathcal{N}$, (ii) all communications between the BS and cellular users occur at the same rate, say $R_s$, i.e., $R_{s,i}=R_s$ $\forall i\in \mathcal{N}$, and (iii) the power consumption of the same type of communications is the same across all cellular users, i.e., $P_{s,i}(R_s)=P_s$, $P_{Tx,i,j}(R_{D2D})=P_{Tx}$ and $P_{Rx,i,j}(R_{D2D})=P_{Rx}$ $\forall i,j\in \mathcal{N}$.  Also, we assume that $R_s$ is much smaller than $R_{D2D}$; specifically, we assume that:
\begin{equation}
\label{EQ:special:case:condition}
\frac{R_s}{R_{D2D}}<\frac{P_s}{P_{Rx}+P_{Tx}}.
\end{equation}
 For instance, such a scenario would occur in practice when  all the cellular users in $\mathcal{N}$ are located far away from the BS but close to each other, \emph{e.g.}, in a stadium or concert hall, and hence data exchange between a pair of users $i$ and $j$ can occur at a fixed and high rate, $R_{D2D}$, and BS-cellular user communication occurs at a lower data rate $R_s$. Fig.~\ref{figure:core} illustrates such a scenario. Finally, we assume that every cellular user in $\mathcal{N}$ has sufficient energy available for relaying services, \emph{i.e.}, $E_i$ is high $\forall i \in \mathcal{N}$.  In this special case game, we will show that \emph{the core is always non-empty}.

 A coalitional game with transferable payoffs $(\mathcal{N},v)$ is \emph{convex} if~\cite{RF:Osborne}:
\begin{equation}
\label{EQ:coalitional:game:convexity}
v(S_1)+v(S_2)\leq v(S_1\cup S_2)+v(S_1\cap S_2), \ \forall S_1,S_2 \subseteq \mathcal{N}.
\end{equation}
It is known that the core of a coalitional game is non-empty if the game is   convex~\cite{RF:Osborne}. We will show that the above special case coalitional game is convex, from which it will follow that the game has a non-empty core.

Consider a coalition $S$. Note that the total energy cost incurred in transferring all the files requested by users in $S$ can be written as: $C(S) = \sum_{m \in \mathcal{M}} aE^m(S)$ (see~\eqref{eq:cost:file}).  The value of a coalition $S$ can be written as:
\begin{align}
\label{EQ:value:separable}
v(S)=\sum\limits_{m \in \mathcal{M}} \left( \sum\limits_{i\in S} d_{i,m} U_{i,m}-aE^m(S)\right). 
\end{align}
The value function is separable in terms of files, \emph{i.e.}, $v(S)$ is the sum of the values obtained from transferring each of the files $i \in \mathcal{M}$. Also, recall that we have assumed that $E_i$ is large for every $i \in \mathcal{N}$. Hence, if we show that \eqref{EQ:coalitional:game:convexity} holds in the case where there is only one file in $\mathcal{M}$, then from \eqref{EQ:value:separable} it will follow that \eqref{EQ:coalitional:game:convexity} holds when there are an arbitrary number of files in $\mathcal{M}$. So in the rest of this section, we consider the case where there is only one file in $\mathcal{M}$; also, assume without loss of generality that this file is of unit size and that $a = 1$. Let $\bar{S} \subseteq S$  be the set of destination nodes in coalition $S$ that request for the file.  For simplicity of notation, we drop the sub-script $m$ and use $U_i$ instead of $U_{i,m}$.  So the value function can be written as:
\begin{equation}
\label{EQ:vS:one:file}
 v(S) = \sum\limits_{i\in \bar{S}}U_i-C(S).
\end{equation} 

Now, to show that \eqref{EQ:coalitional:game:convexity} holds, we first show that the sums of the $U_j$ terms in $v(S_1)+v(S_2)$ and $v(S_1\cup S_2)+v(S_1\cap S_2)$ are equal. By \eqref{EQ:vS:one:file}, the sum of the $U_j$ terms in $v(S_1\cup S_2)+v(S_1\cap S_2)$ is:
\begin{align*}
&\sum\limits_{i\in \bar{S}_1\cup\bar{S}_2}U_i+\sum\limits_{i\in\bar{S}_1\cap\bar{S}_2}U_i\\
&=\sum\limits_{i\in \bar{S}_1}U_i+\sum\limits_{i\in\bar{S}_2}U_i-\sum\limits_{i\in\bar{S}_1\cap \bar{S}_2}U_i+\sum\limits_{i\in\bar{S}_1\cap\bar{S}_2}U_i\\
&=\sum\limits_{i\in \bar{S}_1}U_i+\sum\limits_{i\in\bar{S}_2}U_i.
\end{align*}
So the sums of the $U_i$ terms are equal in $v(S_1)+v(S_2)$ and $v(S_1\cup S_2)+v(S_1\cap S_2)$. Hence, by \eqref{EQ:vS:one:file}, to show that \eqref{EQ:coalitional:game:convexity} holds, it suffices to show that:
\begin{equation}
\label{EQ:coalitional:game:convexity:sufficient:condition}
C(S_1)+C(S_2)\geq C(S_1\cup S_2)+C(S_1\cap S_2), \ \forall S_1, S_2 \subseteq \mathcal{N}. 
\end{equation}

\begin{lemma}
\label{LM:cost:CS}
If \eqref{EQ:special:case:condition} holds, then for a coalition $S \subseteq \mathcal{N}$:
\begin{align*}
C(S)=
\begin{cases}
0,\hspace{1mm}& \mbox{if $\left|\bar{S}\right|=0$},\\
\frac{P_s}{R_s},\hspace{1mm}& \mbox{if $\left|\bar{S}\right|=1$,}\\
\frac{P_s}{R_s}+\frac{P_{Rx}}{R_{D2D}}(\left|\bar{S}\right|-1)+\frac{P_{Tx}}{R_{D2D}},\hspace{1mm}& \mbox {if $\left|\bar{S}\right|\geq2$}.
\end{cases}
\end{align*}
\end{lemma}
\begin{IEEEproof}
The result follows from the fact that if \eqref{EQ:special:case:condition} holds, then when $\bar{S} \neq \emptyset$, the energy cost of coalition $S$ is minimized when the file is downloaded only once from the BS to one of the users in $\bar{S}$ and it is then multicast to the other users, if any, in $\bar{S}$ over D2D links. 
\end{IEEEproof}

We now state the following theorem, which  proves that in the above special case coalitional game, the core is non-empty.
\begin{theorem}
\label{theorem:imputation}
If \eqref{EQ:special:case:condition} holds, then the above special case coalitional game has a non-empty core.
\end{theorem}
\begin{IEEEproof}
We will show that the game is convex, from which the result will follow.
Recall that if \eqref{EQ:coalitional:game:convexity:sufficient:condition} holds, then \eqref{EQ:coalitional:game:convexity} also holds and hence the game is convex.

We now show that \eqref{EQ:coalitional:game:convexity:sufficient:condition} holds in each of the following mutually exclusive and exhaustive cases. 

\emph{Case 1}: If $\bar{S}_1=\bar{S}_2=\emptyset$, then by Lemma~\ref{LM:cost:CS}, $C(S_1)=C(S_2)=C(S_1\cup S_2)=C(S_1\cap S_2)=0$. So \eqref{EQ:coalitional:game:convexity:sufficient:condition} holds.

\emph{Case 2}: If $\bar{S}_1\neq \emptyset$ and $\bar{S}_2= \emptyset$, then by Lemma~\ref{LM:cost:CS}, $C(S_1)=C(S_1\cup S_2)$ and $C(S_2)=C(S_1\cap S_2)=0$. So \eqref{EQ:coalitional:game:convexity:sufficient:condition} holds.

\emph{Case 3}:  If $\bar{S}_1=\emptyset$ and $\bar{S}_2\neq \emptyset$, then by Lemma~\ref{LM:cost:CS}, $C(S_2)=C(S_1\cup S_2)$ and $C(S_1)=C(S_1\cap S_2)=0$.
So \eqref{EQ:coalitional:game:convexity:sufficient:condition} holds.

\emph{Case 4}: If $\bar{S}_1\neq \emptyset$, $\bar{S}_2\neq \emptyset$ and $\bar{S_1}\cap \bar{S}_2=\emptyset$, then by Lemma~\ref{LM:cost:CS}, $C(S_1\cap S_2)=0$. This case can be further divided into sub-cases: a) $\left|\bar{S}_1\right|=\left|\bar{S}_1\right|=1$, b) $\left|\bar{S}_1\right|=1$, $\left|\bar{S}_2\right|\geq 2$, c) $\left|\bar{S}_1\right|\geq 2$, $\left|\bar{S}_2\right|=1$ and d) $\left|\bar{S}_1\right|\geq 2$, $\left|\bar{S}_2\right|\geq 2$. We will show the result for sub-case 4d. The results for the other sub-cases can be similarly shown. Using Lemma \ref{LM:cost:CS}, we get: 
\begin{align}
C(S_1)&=\frac{P_s}{R_s}+\frac{P_{Rx}}{R_{D2D}}(\left|\bar{S}_1\right|-1)+\frac{P_{Tx}}{R_{D2D}},\label{EQ:case4d:CS1} \\
C(S_2)&=\frac{P_s}{R_s}+\frac{P_{Rx}}{R_{D2D}}(\left|\bar{S}_2\right|-1)+\frac{P_{Tx}}{R_{D2D}}, \label{EQ:case4d:CS2}\\ 
C({S}_1\cup {S}_2)&=\frac{P_s}{R_s}+\frac{P_{Rx}}{R_{D2D}}(\left|\bar{S}_1\right|+\left|\bar{S}_2\right|-1)+\frac{P_{Tx}}{R_{D2D}}. \nonumber
\end{align}
So, 
\begin{align*}
C(S_1)+C(S_2)&=2\frac{P_s}{R_s}+\frac{P_{Rx}}{R_{D2D}}(\left|\bar{S}_1\right|+\left|\bar{S}_2\right|-2)+2\frac{P_{Tx}}{R_{D2D}}\\
&=C(S_1\cup S_2)+\frac{P_s}{R_s}+\frac{P_{Tx}}{R_{D2D}}-\frac{P_{Rx}}{R_{D2D}}\\
&> C(S_1\cup S_2).
\end{align*}
The last inequality follows from \eqref{EQ:special:case:condition}. In all the other sub-cases, a similar result can be obtained using Lemma~\ref{LM:cost:CS}. So \eqref{EQ:coalitional:game:convexity:sufficient:condition} holds in Case 4. 

\emph{Case 5}:  $\bar{S}_1\neq \emptyset$, $\bar{S}_2\neq \emptyset$ and $\bar{S_1}\cap \bar{S}_2\neq\emptyset$. This case can also be further divided into sub-cases: a) $\left|\bar{S}_1\right|=\left|\bar{S}_2\right|=\left|\bar{S}_1\cap\bar{S}_2\right|=1$, b) $\left|\bar{S}_1\right|=1$, $\left|\bar{S}_2\right|\geq 2$, $\left|\bar{S}_1\cap\bar{S}_2\right|=1$, c) $\left|\bar{S}_1\right|\geq 2$, $\left|\bar{S}_2\right|=1$, $\left|\bar{S}_1\cap\bar{S}_2\right|=1$, d) $\left|\bar{S}_1\right|\geq 2$, $\left|\bar{S}_2\right|\geq 2$, $\left|\bar{S}_1\cap\bar{S}_2\right|=1$ and e) $\left|\bar{S}_1\right|\geq 2$, $\left|\bar{S}_2\right|\geq 2$, $\left|\bar{S}_1\cap\bar{S}_2\right|\geq 2$. We will show the results for sub-cases 5d and 5e. The results for the other sub-cases can be shown similarly. Consider sub-case 5d. Equations \eqref{EQ:case4d:CS1} and \eqref{EQ:case4d:CS2} hold in this sub-case.   Also: 
\begin{align*}
C({S}_1\cup {S}_2)&=\frac{P_s}{R_s}+\frac{P_{Rx}}{R_{D2D}}(\left|\bar{S}_1\cup \bar{S}_2\right|-1)+\frac{P_{Tx}}{R_{D2D}}\\
&= \frac{P_s}{R_s}+\frac{P_{Rx}}{R_{D2D}}(\left|\bar{S}_1\right|+\left| \bar{S}_2\right|-2)+\frac{P_{Tx}}{R_{D2D}}\\
C(S_1\cap S_2)&=\frac{P_s}{R_s}
\end{align*}
The last two equalities hold since $\left|\bar{S}_1\cap \bar{S}_2\right|=1$. It can be easily seen that $C(S_1)+C(S_2)> C(S_1\cup S_2)+C(S_1\cap S_2)$. Now consider sub-case 5e. Equations \eqref{EQ:case4d:CS1} and \eqref{EQ:case4d:CS2} hold in this sub-case. Also:
\begin{align*}
C({S}_1\cup {S}_2)&=\frac{P_s}{R_s}+\frac{P_{Rx}}{R_{D2D}}(\left|\bar{S}_1\cup \bar{S}_2\right|-1)+\frac{P_{Tx}}{R_{D2D}}\\
&= \frac{P_s}{R_s}+\frac{P_{Rx}}{R_{D2D}}(\left|\bar{S}_1\right|+\left| \bar{S}_2\right|-\left|\bar{S}_1\cap \bar{S}_2\right|-1)\\&\quad+\frac{P_{Tx}}{R_{D2D}}\\
C(S_1\cap S_2)&=\frac{P_s}{R_s}+\frac{P_{Rx}}{R_{D2D}}(\left|\bar{S}_1\cap \bar{S}_2\right|-1)+\frac{P_{Tx}}{R_{D2D}}
\end{align*}
From the above, it can be easily seen that $C(S_1)+C(S_2)=C(S_1\cup S_2)+C(S_1\cap S_2)$. Similarly, using Lemma~\ref{LM:cost:CS}, it can be easily checked that in sub-cases 5a, 5b and 5c, we have $C(S_1)+C(S_2)=C(S_1\cup S_2)+C(S_1\cap S_2)$. So \eqref{EQ:coalitional:game:convexity:sufficient:condition} holds in Case 5.

The result follows.
\end{IEEEproof}

Theorem~\ref{theorem:imputation} shows that although, in general, the above coalitional game under Model $A$ may have an empty core (see Example~\ref{EG:empty:core}), in the special case game wherein all communication links are symmetric across the cellular users and $R_s$ is much smaller than $R_{D2D}$, the core is always non-empty. Intuitively, this is because multiple clusters such as $S_1$ and $S_2$ as in Example~\ref{EG:empty:core}, such that the achievable data rate between user $i_1$ and user $i_2$ is low  for every pair $i_1 \in S_1$ and $i_2 \in S_2$ and the achievable data rates among users within each cluster are high, cannot exist in the special case game due to the fact that the achievable data rates between different pairs of cellular users are equal; also, since $R_s$ is much smaller than $R_{D2D}$, the energy required to download a file from the BS only once and multicast it over D2D links is less than that required when it is downloaded from the BS multiple times. So when all the users in $\mathcal{N}$ cooperate, a smaller amount of energy is required for the file transfer than when they do not cooperate. Hence, it is beneficial for all the users in the grand coalition $\mathcal{N}$ to cooperate.

\section{Coalition Formation Under Model A}
\label{SC:coalition:formation}
In the previous section, we provided sufficient conditions under which the coalitional game with transferable payoffs $(v,\mathcal{N})$ has a non-empty core. However, when these conditions are not satisfied, it may not be beneficial for \emph{all} the users to cooperate among themselves. So in this section, we seek conditions under which the set of cellular users can be partitioned into groups
(coalitions) such that it is beneficial for the users of each coalition to cooperate among themselves. For this purpose, we use the framework of stable partitions defined in~\cite{RF:stable:coalitions}, a brief summary of which is presented in Section~\ref{SSC:Dc:stable:preliminaries}.

\subsection{Mathematical Preliminaries}
\label{SSC:Dc:stable:preliminaries}
\ignore{We now review the framework of stable partitions described in~\cite{RF:stable:coalitions}.}
A \emph{collection} $\mathbf{S}=\{S_1,\ldots,S_k\}$ in the set $\mathcal{N}$ is a set of mutually disjoint coalitions of the users in the set $\mathcal{N}$. A set $\mathbf{P}=\{P_1,\ldots,P_n\}$ is called a \emph{partition} of the set $\mathcal{N}$, if $P_i\subseteq \mathcal{N}$, $P_i\cap P_j=\phi$ for all $P_i,P_j\in\mathbf{P}$ and $\cup_{i=1}^nP_{i}=\mathcal{N}$.  Consider a partition $\mathbf{P}=\{P_1,\ldots,P_n\}$ and a collection of coalitions $\mathbf{S}=\{S_1,\ldots,S_k\}$. The term $\mathbf{S}\left[\mathbf{P}\right]$ is defined as follows:
\begin{equation}
\mathbf{S}\left[\mathbf{P}\right]=\{\cup_{i=1}^k S_i\cap P_1,\ldots,\cup_{i=1}^k S_i\cap P_n\}.
\end{equation}  
$\mathbf{S}\left[\mathbf{P}\right]$ is the partitioning of users in $\cup_{i=1}^k S_i$ into coalitions according to the partition $\mathbf{P}$. For every collection of coalitions $\mathbf{S}=\{S_1,\ldots,S_k\}$, let:
\begin{equation}
v(\mathbf{S})=\sum\limits_{i=1}^kv(S_i)
\end{equation}
denote the \emph{value} of collection $\mathbf{S}$. A collection $\mathbf{S}=\{S_1,\ldots,S_k\}$ is \emph{$\mathbf{P}-$compatible} if $\cup_{i=1}^k S_i\subseteq P_j$ for some coalition $P_j\in \mathbf{P}$ and a coalition $S$ is \emph{$\mathbf{P}-$incompatible} if $S\not\subset P_i$ for every $P_i\in\mathbf{P}$.  In this work, we use the concept of \emph{$\mathbb{D}_c$-stability.} 
\begin{definition}
A partition $\mathbf{P}$ is $\mathbb{D}_c$-stable if and only if:
\begin{equation}
\label{eq:stable:condition}
v(\mathbf{S}\left[\mathbf{P}\right])\geq v(\mathbf{S}) \hspace{2mm}
\end{equation}
for every collection $\mathbf{S}$ in $\mathcal{N}$~\cite{RF:stable:coalitions}.
\end{definition}

The value of a coalition $S$ can be interpreted as the excess that is available to be distributed among the members of the coalition $S$. The above definition says that a partition $\mathbf{P}$ is $\mathbb{D}_c$-stable if, for every collection $\mathbf{S}=\{S_1,\ldots,S_k\}$ in $\mathcal{N}$, the excess available to users in $\cup_{i=1}^k S_i$ when they partition according to $\mathbf{S}$ is at most the excess available when these users form coalitions according to the partition $\mathbf{P}$. The following result provides a useful necessary and sufficient condition for a partition to be $\mathbb{D}_c$-stable.
\begin{theorem}
\label{theorem:stable:conditions}
A partition $\mathbf{P} = \{P_1,\ldots,P_n\}$ of $\mathcal{N}$ is $\mathbb{D}_c$-stable if and only if the following two conditions are satisfied~\cite{RF:stable:coalitions}:
\begin{enumerate}
\item
For every $\mathbf{P}-$compatible collection $\mathbf{S}=\{S_1,\ldots,S_k\}$ we have:
\begin{equation}
\label{eq:stable:p:compatible}
v(\cup_{i=1}^k S_i)\geq\sum\limits_{i=1}^kv(S_i).
\end{equation}
\item
For every $\mathbf{P}-$incompatible coalition $S$, we have:
\begin{equation}
\label{eq:stable:p:incompatible}
\sum\limits_{i=1}^nv(S\cap P_{i})\geq v(S).
\end{equation}
\end{enumerate}
\end{theorem}

The above conditions can be interpreted as follows: consider a $\mathbb{D}_c$-stable partition $\mathbf{P}=\{P_1,\ldots,P_n\}$. Inequality \eqref{eq:stable:p:compatible} says that the users of every subset of a coalition $P_k\in\mathbf{P}$ are better off forming a coalition among themselves rather than dividing themselves into multiple coalitions and \eqref{eq:stable:p:incompatible} says that users of different coalitions $P_1,\ldots,P_n$ are better off splitting according to partition $\mathbf{P}$ than forming a $\mathbf{P}$-incompatible coalition. Further, if the inequalities in \eqref{eq:stable:p:compatible} and \eqref{eq:stable:p:incompatible} hold with strict inequalities, then
 partition $\mathbf{P} = \{P_1,\ldots,P_n\}$  is said to be \emph{strictly $\mathbb{D}_c$-stable .} A special characteristic of a $\mathbb{D}_c$-stable partition is stated in the following proposition. 

\begin{proposition}
\label{prop:stable:highest:value}
If $\mathbf{P}$ is a $\mathbb{D}_c$-stable partition of $\mathcal{N}$, then $v(\mathbf{P})=\max\{v(\mathbf{Q}): \mbox{$\mathbf{Q}$ is a partition of $\mathcal{N}$} \}$~\cite{RF:stable:coalitions}.
\end{proposition}

To find the $\mathbb{D}_c$-stable partition of a given set, we employ the \emph{merge and split algorithm}~\cite{RF:stable:coalitions}. The algorithm takes an arbitrary partition as its input and repeatedly performs merge and split operations on the coalitions of the partition, whenever certain conditions are satisfied, until these operations are no longer possible. We now state the conditions under which coalitions can be merged or split. Consider a partition $\mathbf{I}=\{M_1,\ldots,M_p,S_1,\ldots,S_q\}$. 
\begin{enumerate}
\item
The coalitions $M_1,\ldots,M_p$ can be merged to form a partition $\mathbf{I_1}=\{\cup_{i=1}^pM_i,S_1,\ldots,S_q\}$ if $\sum\limits_{i=1}^pv(M_i)<v(\cup_{i=1}^pM_i)$. 
\item
A coalition $S_i$ can be split into $S_{i,1},\ldots,S_{i,n}$ to form a partition $\mathbf{I_2}=\{M_1,\ldots,M_p,S_1,\ldots,S_{i-1},S_{i,1},\ldots,\\S_{i,n},S_{i+1},\ldots,S_q\}$ if $v(S_i)< \sum\limits_{k=1}^{n}v(S_{i,k})$. 
\end{enumerate}

In the merge and split algorithm, we start from an arbitrary partition and repeatedly perform the merge and split operations in any order   
until no merging or splitting operation is possible. When the algorithm terminates, we get a partition such that the merging of two or more coalitions or splitting of a coalition of the partition does not strictly increase the value of the partition. 

\begin{proposition}
\label{prop:merge:split}
The above merge and split algorithm terminates after a finite number of operations~\cite{RF:stable:coalitions}. 
\end{proposition}

\begin{theorem}
\label{theorem:merge:and:split}
Suppose $\mathbf{P}$ is a strictly $\mathbb{D}_c$-stable partition of $\mathcal{N}$. Then the merge and split algorithm starting from an arbitrary partition of $\mathcal{N}$ converges to $\mathbf{P}$. Also, $\mathbf{P}$ is the unique $\mathbb{D}_c$-stable partition~\cite{RF:stable:coalitions}.
\end{theorem}

\subsection{$\mathbb{D}_c$-Stability of a Partition}
\begin{center}
\begin{figure}[!hbt]
\centering
\includegraphics[scale=0.23]{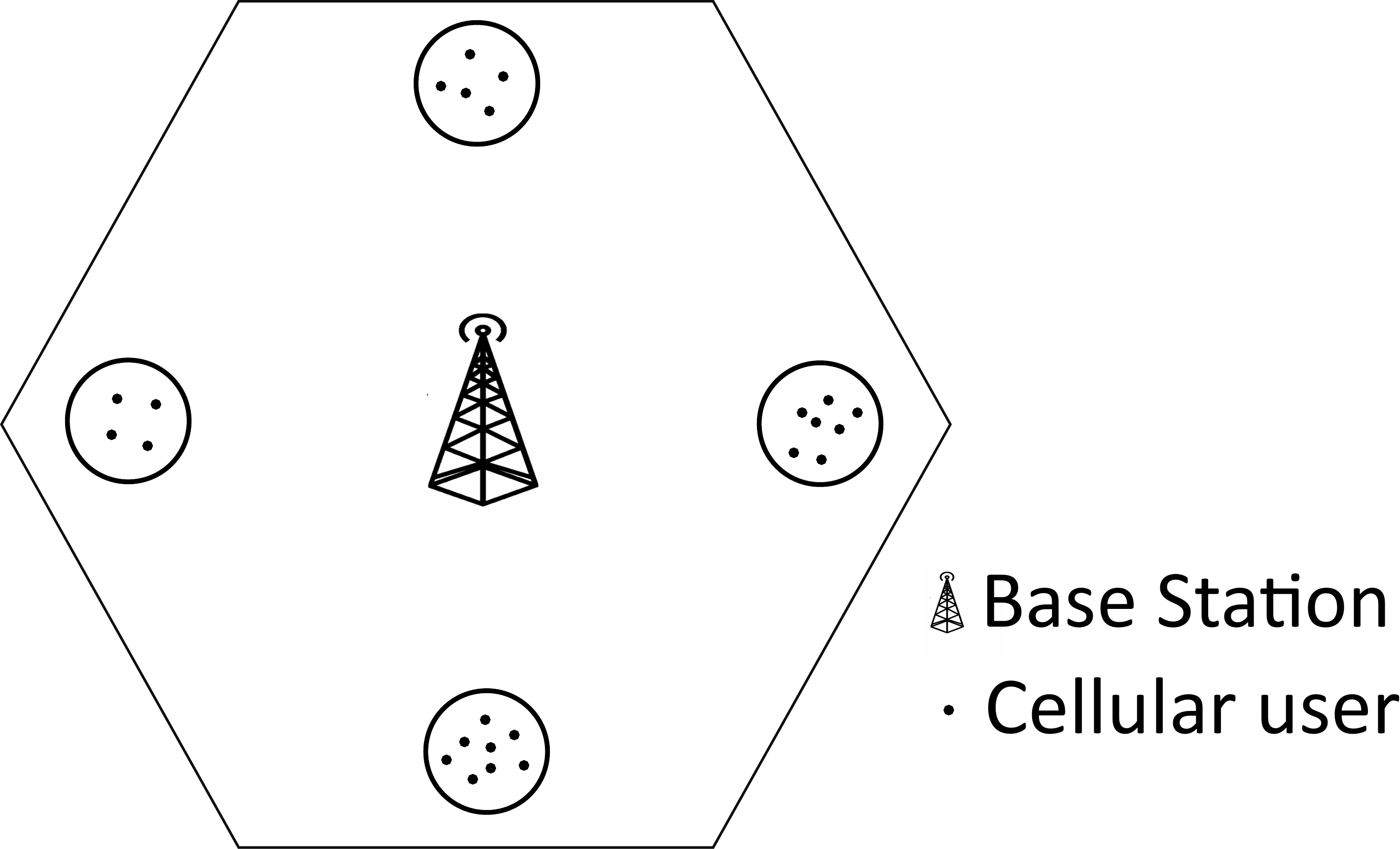}
\caption{The figure shows a single cell with a BS at the center and cellular users located in four clusters represented by circles. Each cluster is located far away from every other cluster, but cellular users belonging to a given cluster are located close to each other.}
\label{figure:clusters}
\end{figure}
\end{center}

In general, it is not necessary for a $\mathbb{D}_c$-stable partition to exist. So in this section, for the coalitional game under Model A, we find a set of sufficient conditions under which a given partition $\mathbf{P}=\{P_1,\ldots,P_n\}$ is $\mathbb{D}_c$-stable. Note that from Proposition~\ref{prop:stable:highest:value}, a $\mathbb{D}_c$-stable partition has the highest value among all possible partitions of $\mathcal{N}$. Also, note that a partition $\mathbf{P}$ is $\mathbb{D}_c$-stable if and only if \eqref{eq:stable:p:compatible} and \eqref{eq:stable:p:incompatible} are satisfied. We consider a cell wherein cellular users are located in multiple clusters $P_1,\ldots,P_n$ such that the users of each cluster are located close to each other and users of different clusters are located far away from each other (see Fig.~\ref{figure:clusters}). We investigate conditions under which such a partition $\mathbf{P}=\{P_1,\ldots,P_n\}$ is $\mathbb{D}_c$-stable. 

Since users in a given cluster are located close to each other, we assume that users in a given cluster have symmetric connections to the BS and symmetric D2D connections among themselves. We also assume that D2D connections are symmetric across users of different coalitions. Formally, when cellular users $i,j\in P_k$, we have: 
\begin{subequations}
\label{eq:same:cluster}
\begin{align}
R_{s,i}&=R_{s,P_k}\label{eq:same:cluster:data:rate},\\
P_{s,i}(R_{s,i})&=P_{s,P_k}\label{eq:same:cluster:power},\\
R_{D2D,i,j}&=R_{D2D,P_k}\label{eq:same:cluster:D2D},\\
P_{Tx,i,j}(R_{D2D,P_k})&=P_{Tx,P_k}\label{eq:same:cluster:D2D:tx:power},\\
P_{Rx,i,j}(R_{D2D,P_k})&=P_{Rx,P_k}\label{eq:same:cluster:D2D:rx:power},
\end{align}
\end{subequations}
and when $i\in P_k,j\in P_l$, we have:
\begin{subequations}
\label{eq:different:cluster}
\begin{align}
R_{D2D,i,j}&=R_{D2D,P_k,P_l}\label{eq:different:cluster:D2D:rate},\\
P_{Tx,i,j}(R_{D2D,P_k,P_l})&=P_{Tx,P_k,P_l}\label{eq:different:cluster:D2D:tx:power},\\
P_{Rx,i,j}(R_{D2D,P_k,P_l})&=P_{Rx,P_k,P_l}\label{eq:different:cluster:D2D:rx:power}.
\end{align}
\end{subequations}

Let $S$ represent an arbitrary $\mathbf{P}$-incompatible coalition and for every $i\in S$, let $P^i\in \mathbf{P}$ represent the cluster of user $i$. Let $i$ and $j$ be such that $P^i\neq P^j$. We define the following notations:
\begin{subequations}
\label{eq:notations}
\begin{center}
\begin{equation}
E_{min,S}^s=\min\limits_{i\in S}\left\{\frac{P_{s,i}}{R_{s,i}}\right\},
\end{equation}
\begin{equation}
E_{max,S}^s=\max\limits_{i\in S}\left\{\frac{P_{s,i}}{R_{s,i}}\right\},
\end{equation}
\begin{equation}
E_{min,S}^t=\min\limits_{i,j\in S, P^i\neq P^j}\left\{\frac{P_{Tx,P^i,P^j}}{R_{D2D,P^i,P^j}}\right\},
\end{equation}
\begin{equation}
E_{min,S}^r=\min\limits_{i,j\in S, P^i\neq P^j}\left\{\frac{P_{Rx,P^i,P^j}}{R_{D2D,P^i,P^j}}\right\},
\end{equation}
\begin{equation}
E_{max}^{D2D}=\max\limits_{P_k\in \mathbf{P}}\frac{P_{Rx,P_k}+P_{Tx,P_k}}{R_{D2D,P_k}},
\end{equation}
\begin{equation}
R_{D2D,i,S}=\min _{j\in S}R_{D2D,i,j}.
\end{equation}
\end{center}
\end{subequations}
Also, $P_{Tx,i,S}(R_{D2D,i,S})$ is the power required to multicast a file by user $i$ to users in coalition $S$ at the rate $R_{D2D,i,S}$. 

Now, the value of coalition $S$ is given by the maximum value of the objective function in problem \eqref{EQ:PS}. As shown in Section~\ref{SSC:model:A:special}, the value of coalition $S$ is separable in terms of files (see \eqref{EQ:value:separable}). So if $E_j$ is assumed to be large enough and we show that \eqref{eq:stable:p:compatible}, \eqref{eq:stable:p:incompatible} are satisfied in
the case where there is only one file in $\mathcal{M}$, then from \eqref{EQ:value:separable}, it will follow that \eqref{eq:stable:p:compatible}, \eqref{eq:stable:p:incompatible} are satisfied in
the case where there are an arbitrary number of files in $\mathcal{M}$, and hence that in the latter case, the partition $\mathbf{P}$ is $\mathbb{D}_c$-stable. So from this point, we assume that all the requests from the cellular users in $S$ are for a single file. We also assume, without loss of generality, that the file is of unit size and $a=1$. It can be easily seen that the valuation ($U_{i,j}$) terms in $v(\cup_{i=1}^k S_i)$ and $\sum\limits_{i=1}^kv(S_i)$  (respectively, $v(S)$ and $\sum\limits_{i=1}^n v(S\cap P_i)$), which appear in \eqref{eq:stable:p:compatible} (respectively, \eqref{eq:stable:p:incompatible}), are the same. So henceforth, while studying sufficient conditions for \eqref{eq:stable:p:compatible} and  \eqref{eq:stable:p:incompatible} to hold,  we only consider the cost terms (hence, energy terms). 

We now state the main result of this section. 
\begin{theorem}
\label{theorem:main:result}
A partition $\mathbf{P}=\{P_1,\ldots,P_n\}$ of $\mathcal{N}$ satisfying~\eqref{eq:same:cluster} and  \eqref{eq:different:cluster} is $\mathbb{D}_c$-stable if the following three conditions are satisfied:
\begin{enumerate}
\item
$E_{max,\mathcal{N}}^s\leq \min\limits_{k\in\{2,\ldots,n\}} \left( \frac{E_{min,\mathcal{N}}^s}{k}+\frac{E_{min,\mathcal{N}}^t}{k}+\frac{k-1}{k}E_{min,\mathcal{N}}^r \right)$,
\item
$E_{max}^{D2D}\leq E_{min,\mathcal{N}}^r$,
\item 
$\min\limits_{P_k\in \mathbf{P}} \left\{\frac{P_{s,P_k}}{R_{s,P_k}}\right\}> \max\limits_{P_k\in \mathbf{P}}\left\{\frac{P_{Tx,P_k}+P_{Rx,P_k}}{R_{D2D,P_k}}\right\}.$
\end{enumerate}
\end{theorem}

The significance of Theorem~\ref{theorem:main:result} is as follows. By Proposition~\ref{prop:stable:highest:value}, out of the partitions of $\mathcal{N}$, a $\mathbb{D}_c$-stable partition is the partition with the maximum possible value. Since the sum of the valuation ($U_{i,j}$)  terms in the value of  each partition is the same (see \eqref{EQ:value:separable}), the \emph{$\mathbb{D}_c$-stable partition is the partition in which the minimum total energy is expended by users during the transfer of files from the BS to the destination nodes.} Thus, Theorem~\ref{theorem:main:result} provides a set of sufficient conditions for a partition to be the one in which the minimum total energy is expended by users.


The conditions in Theorem~\ref{theorem:main:result} can be interpreted as follows:
\begin{itemize}
\item
Condition 1)  states that for every $k \in \{2, \ldots, n\}$, the following inequality holds:
\begin{equation}
k E_{max,\mathcal{N}}^s \leq E_{min,\mathcal{N}}^s+E_{min,\mathcal{N}}^t + (k-1) E_{min,\mathcal{N}}^r. 
\label{EQ:thm:1:cond:2:fixed:k}
\end{equation}
The LHS of \eqref{EQ:thm:1:cond:2:fixed:k} represents the maximum energy consumed when there are $k$ cellular users and all of them download their file directly from the BS. The RHS represents the minimum energy consumed when the file is downloaded from the BS only once by a user, say $i$, and multicast over D2D links to $(k-1)$ users, each of which belongs to a coalition different from $P^i$, which is the coalition to which $i$ belongs. Thus, Condition 1) says that less energy is expended when users of different clusters download files directly from the BS than when they cooperate among themselves and distribute the file over D2D links.
\item
Condition 2) says that the energy consumed in transmission and reception of a file over the D2D link between two users, both of which are in the same coalition, say $P_k$, is less than or equal to the energy consumed by a user in downloading the file from the user of a different coalition. 
\item
Condition 3) says that the energy required for any user to download a file of unit size from the BS is greater than the total energy consumed in transmission and reception during the download of a file of unit size by any user (say, $i$) from a relay of its own cluster ($P^i$) over a D2D link.
\end{itemize}

Now, we state and prove two lemmas (Lemmas~\ref{lemma:p:compatible} and~\ref{lemma:p:incompatible}), from which the proof of Theorem~\ref{theorem:main:result} follows.
\begin{lemma}
\label{lemma:p:compatible}
If  the equalities in~\eqref{eq:same:cluster} are satisfied for every coalition $P_k\in\mathbf{P}$ and Condition 3) in Theorem~\ref{theorem:main:result} holds, then \eqref{eq:stable:p:compatible} holds for every $\mathbf{P}-$compatible collection $\mathbf{S}$.
\end{lemma}
\begin{IEEEproof}
Consider the coalitional game with transferable payoffs $(P_k,v)$, $P_k\in\mathbf{P}$.  If $v$ satisfies the superadditivity property\footnote{A coalition game with transferable payoffs $(v,\mathcal{N})$ is \emph{superadditive} if for every $S_1,S_2\subseteq \mathcal{N}$ where $S_1\cap S_2=\emptyset$, we have $v(S_1\cup S_2)\geq v(S_1)+v(S_2)$~\cite{RF:Osborne}.} for each of the coalitional games $(P_k, v)$, $P_k\in\mathbf{P}$, then \eqref{eq:stable:p:compatible} is satisfied. 
Also, recall from the proof of Theorem~\ref{theorem:imputation} that if:
\begin{equation*}
\label{eq:condition:1}
\frac{R_{s,P_k}}{R_{D2D,P_k}}<\frac{P_{s,P_k}}{P_{Rx,P_k}+P_{Tx,P_k}},
\end{equation*}
then the coalitional game with transferable payoffs $(P_k,v)$ is a convex game. The result follows from the fact that a convex game satisfies the superadditivity property \cite{RF:Osborne}. 
\end{IEEEproof}

\begin{lemma}
\label{lemma:p:incompatible}
If Conditions 1), 2), 3) in Theorem~\ref{theorem:main:result} along with \eqref{eq:same:cluster},~\eqref{eq:different:cluster} are statisfied, then \eqref{eq:stable:p:incompatible} holds for every $\mathbf{P}-$incompatible coalition $S$.
\end{lemma}
\begin{IEEEproof}
Recall that we assume that all the requests from the cellular users in $S$ are for a single file. Let $\bar{S} \subseteq S$ denote the destination nodes that request the file from the BS.

We prove the result for the following two mutually exclusive and exhaustive cases:

Case 1): $\bar{S}$ is a $\mathbf{P}$-compatible coalition. Consider the following sub-cases:
\begin{itemize}
\item[a.]
$\left|\bar{S}\right|=1$. Suppose cellular user $l$ requests the file. From \eqref{eq:cost:file}, we can write (ignoring superscript $m$):
\begin{align}
\label{eq:coalition:cost}
E(S)=\sum\limits_{i\in S}\frac{\alpha_iP_{s,i}}{R_{s,i}}+\sum\limits_{i\in S\setminus\{l\}}\frac{\alpha_i P_{Tx,i,l}}{R_{D2D,i,l}}\nonumber\\+\sum\limits_{j\in S\setminus\{l\}}\frac{\alpha_jP_{Rx,j,l}}{R_{D2D,j,l}}. 
\end{align}
If $\alpha_l=1$, then user $l$ downloads the entire file directly from the BS. If $0\leq \alpha_l<1$, then a part of the file is downloaded by other users and transmitted to user $l$. The coalition $S$ may contain users from coalition $P^l$. Since all the users in coalition $P^l$ can download from the BS at the same rate and reception power, $E(S)$ is minimised when $\alpha_i=0$ for all $i\in S\cap P^l\setminus \{l\}$. When $i\in S\setminus P^l$, the total cost of transferring $\alpha_i$ fraction of the file from the BS to user $l$ is $\frac{\alpha_{i}P_{s,i}}{R_{s,i}}+\frac{\alpha_iP_{Tx,i,l}}{R_{D2D,i,l}}+\frac{\alpha_iP_{Rx,i,l}}{R_{D2D,i,l}}$  which is greater than $\frac{\alpha_iP_{s,l}}{R_{s,l}}$ (see Condition 1) of Theorem~\ref{theorem:main:result}).
\item[b.]
$\left|\bar{S}\right|>1$ and $\bar{S}\subseteq P_l$ for some $P_l\in\mathbf{P}$. Let $\alpha_{P}=\sum_{i\in P}\alpha_i$. We have, from \eqref{eq:cost:file}, the following: 
\begin{align}
&E(S)=\nonumber\\
&\sum\limits_{i\in S}\frac{\alpha_i P_{s,i}}{R_{s,i}}+\sum\limits_{i\in S}\frac{\alpha_iP_{Tx,i,\bar{S}\setminus\{i\}}}{R_{D2D,i,\bar{S}\setminus\{i\}}}+\sum\limits_{i\in\bar{S}}\sum\limits_{j\neq i}\frac{\alpha_j P_{Rx,j,i}}{R_{D2D,j,i}}\nonumber
\end{align}
By separating the terms corresponding to $i\in S\cap P_l$ and $i\in S\setminus P_l$, we get:
\begin{align}
&E(S)=\nonumber\\
&\sum\limits_{i\in S\cap P_l}\Bigg(\frac{\alpha_iP_{s,i}}{R_{s,i}}+\frac{\alpha_iP_{Tx,i,\bar{S}\setminus\{i\}}}{R_{D2D,i,\bar{S}\setminus\{i\}}}\Bigg)+\sum\limits_{i\in S\setminus P_l}\Bigg(\frac{\alpha_iP_{s,i}}{R_{s,i}}+\nonumber\\&\quad\frac{\alpha_iP_{Tx,i,\bar{S}}}{R_{D2D,i,\bar{S}}}\Bigg)+\sum\limits_{i\in \bar{S}}\Bigg(\sum\limits_{j\neq i,j\in S\cap P_l}\frac{\alpha_jP_{Rx,j,i}}{R_{D2D,j,i}}+\nonumber\\&\quad\sum\limits_{j\in S\setminus P_l}\frac{\alpha_jP_{Rx,j,i}}{R_{D2D,j,i}}\Bigg)\nonumber\\
&\geq \alpha_{S\cap P_l}\Bigg(\frac{P_{s,P_l}}{R_{s,P_l}}+\frac{P_{Tx,P_l}}{R_{D2D,P_l}}+\left|\bar{S}\right|\frac{P_{Rx,P_l}}{R_{D2D,P_l}}\Bigg)\nonumber\\&\quad-\alpha_{\bar{S}}\frac{P_{Rx,P_l}}{R_{D2D,P_l}}\nonumber\\
&\quad+\alpha_{S\setminus P_l}\Big(E_{min,S\setminus P_l}^s+E_{min,S\setminus P_l}^t+\left|\bar{S}\right|E_{min,S\setminus P_l}^r\Big)\nonumber\\
&\geq \alpha_{S\cap P_l}\Bigg(\frac{P_{s,P_l}}{R_{s,P_l}}+\frac{P_{Tx,P_l}}{R_{D2D,P_l}}+(\left|\bar{S}\right|-1)\frac{P_{Rx,P_l}}{R_{D2D,P_l}}\Bigg)\nonumber\\
&\quad+\alpha_{S\setminus P_l}(E_{max,\mathcal{N}}^s+(\left|\bar{S}\right|-1)E_{max}^{D2D})\nonumber\\
&\quad+(\alpha_{S\cap P_l}-\alpha_{\bar{S}})\frac{P_{Rx,P_l}}{R_{D2D,P_l}}\nonumber\\
&\geq E(S\cap P_l)\nonumber.
\end{align}
We get the first inequality by lower bounding the terms $\frac{P_{s,i}}{R_{s,i}}$, $\frac{P_{Tx,i,\bar{S}}}{R_{D2D,i,\bar{S}}}$ and $\frac{P_{Rx,j,i}}{R_{D2D,j,i}}$ for $i\in S\setminus P_l$  by  $E_{min,S\setminus P_l}^s$, $E_{min,S\setminus P_l}^t$ and $E_{min,S\setminus P_l}^r$ respectively. The second inequality follows from Conditions 1) and 2) of Theorem \ref{theorem:main:result}. The third inequality follows from the definitions of $E_{max,\mathcal{N}}^s$ and $E_{max}^{D2D}$ (see \eqref{eq:notations}).
\end{itemize}

Case 2): $\bar{S}$ is a $\mathbf{P}$-incompatible coalition. Note that $\left|\bar{S}\right|>1$. Consider the following sub-cases:
\begin{itemize}
\item[a.]
$2\leq \left|\bar{S}\right|\leq n$. Then,
\begin{align}
\label{eq:EC:inequality}
&E(S)\nonumber\\
&=\sum\limits_{i\in S}\frac{\alpha_i P_{s,i}}{R_{s,i}}+\sum\limits_{i\in S}\frac{\alpha_iP_{Tx,i,\bar{S}\setminus\{i\}}}{R_{D2D,i,\bar{S}\setminus\{i\}}}+\sum\limits_{i\in\bar{S}}\sum\limits_{j\neq i}\frac{\alpha_j P_{Rx,j,i}}{R_{D2D,j,i}}\nonumber\\
&\geq E_{min,S}^s\sum\limits_{i\in S}\alpha_i+E_{min,S}^t\sum\limits_{i\in S}\alpha_i +E_{min,S}^r\sum\limits_{i\in\bar{S}}\sum\limits_{j\neq i}\alpha_j\nonumber\\
&\geq E_{min,S}^s+E_{min,S}^t+(\left|\bar{S}\right|-1)E_{min,\bar{S}}^r\\
& =  \left|\bar{S}\right|\Bigg(\frac{E_{min,S}^s}{\left|\bar{S}\right|}+\frac{E_{min,S}^t}{\left|\bar{S}\right|}+\frac{\left|\bar{S}\right|-1}{\left|\bar{S}\right|}E_{min,S}^r\Bigg)\nonumber\\
&\geq \left|\bar{S}\right|E_{max,S}^s\nonumber\\
&\geq \sum\limits_{i=1}^n E(S\cap P_i)\nonumber
\end{align}
The first inequality follows from \eqref{eq:notations}. The third inequality follows from Condition 1) of Theorem~\ref{theorem:main:result}. The fourth inequality follows from Lemma~\ref{lemma:p:compatible}, which implies that users consume less energy when they download files from relays of their own clusters over D2D links rather than downloading them directly from the BS.
\item[b.]
$\left|\bar{S}\right|>n$. The lower bound for $E(S)$ in \eqref{eq:EC:inequality} still holds. \ignore{and the term $E(S)$ is maximum when at least one user from every coalition $P_l\in \mathbf{P}$ requests a file and all other requests come from the coalition corresponding to $E_{max}^{D2D}$. We can upper bound the energy consumed in downloading the file by each of the $n$ users directly from the BS by $E_{max,\mathcal{N}}^s$.} Thus we get:
\begin{align*}
E(S)&\geq E_{min,S}^s+E_{min,S}^t+(|\bar{S}|-1)E_{min,\bar{S}}^r\\
&\geq nE_{max,\mathcal{N}}^s+(|\bar{S}|-n)E_{min,\bar{S}}^r\\
&\geq nE_{max,\mathcal{N}}^s+(|\bar{S}|-n)E_{max}^{D2D}\\
&\geq \sum\limits_{i=1}^n E(S\cap P_i)
\end{align*}
The second inequality follows from Condition 1) of Theorem~\ref{theorem:main:result}, the third inequality follows from Condition 2) of Theorem~\ref{theorem:main:result} and the fourth inequality follows from Lemma~\ref{lemma:p:compatible}.
\end{itemize}
The result follows.
\end{IEEEproof}

Finally, Theorem~\ref{theorem:main:result} follows from Lemmas~\ref{lemma:p:compatible} and~\ref{lemma:p:incompatible} and Theorem~\ref{theorem:stable:conditions}. 

The following proposition provides an algorithm to efficiently compute a $\mathbb{D}_c$-stable partition. 

\begin{proposition}
If a strictly $\mathbb{D}_c$-stable partition exists, then the merge and split algorithm proposed in~\cite{RF:coalition:3} converges to it in polynomial time. Specifically, the time complexity of the algorithm is $O(N^2)$ for our network model.
\end{proposition}

The proof of the above proposition is straightforward and is omitted for brevity.

\section{NP-Completeness of and Heuristics for the Relay Assignment Problem under Model $B$}
\label{SC:model:B}
Consider Model B defined in Section~\ref{SC:system:model}. Recall that in this model, to find the value, $v(S)$, of a coalition $S$, we need to maximize the sum of utilities of all the cellular users in $S$; for this, we in turn need to solve the optimization problem $P(S)$ defined in Section~\ref{SC:system:model} (see \eqref{EQ:PS}) with the constraints  $\alpha_{i,m} \in \{0,1\}$, for all $i \in \mathcal{N}, m \in \mathcal{M}$. We refer to this problem as problem $P_B(S)$. Unfortunately, it turns out that problem $P_B(S)$ is an NP-Complete problem~\cite{RF:kleinberg:algorithm}; we show this NP-Completeness in Section~\ref{SSC:model:B:NP:completeness}. Hence, it is computationally prohibitive to find the value, $v(S)$, of a coalition. So we do not perform a cooperative game theoretic analysis of Model $B$. However, we  provide heuristics to solve problem $P_B(S)$ in Section~\ref{SSC:model:B:heuristics} and evaluate their performance via numerical studies in Section~\ref{SC:numerical:results}.   

\subsection{NP-Completeness}
\label{SSC:model:B:NP:completeness}
Let $\mathcal{M}_S\subseteq \mathcal{M}$ denote the set of all files that are requested by at least one user in $S$.
\begin{theorem}
Problem $P_B(S)$ is NP-Complete. 
\end{theorem}
\begin{IEEEproof}
First, it is easy to check that problem $P_B(S)$ is in class NP~\cite{RF:kleinberg:algorithm}. We now prove the NP-Completeness of problem $P_B(S)$ by reducing the generalized assignment problem (GAP)~\cite{RF:Combinatorial}, which is known to be NP-Complete, to a special case of problem $P_B(S)$.  

The GAP deals with the allotment of jobs to agents. Let $\mathcal{M}_S$ (respectively, $S$) be the set of all jobs (respectively, agents). Agent $i \in S$ incurs a cost $c_{i,m}$ when it performs job $m \in \mathcal{M}_S$ and agent $i$ has a total budget of $t_i$. When job $m$ is assigned to agent $i$, a profit of $p_{i,m}$ is gained. The objective of the GAP is to assign an agent to each job so as to maximize the total profit from all the assignments of agents to jobs, while satisfying the budget constraint of each agent. Let $\alpha_{i,m}$ be $1$ if agent $i$ is assigned to job $m$ and $0$ else. The GAP can be written as:
\begin{center}
$\max\sum\limits_{i,m}p_{i,m}\alpha_{i,m}$
\end{center}
subject to:\\
\noindent
1) $\alpha_{i,m}\in\{0,1\}, \ \forall i \in S, m \in \mathcal{M}_S$, \\
2) $\sum\limits_{i\in S}\alpha_{i,m}=1,  \ \forall m\in \mathcal{M}_S$,\\
3) $\sum\limits_{m\in \mathcal{M}_S}\alpha_{i,m}c_{i,m}\leq t_m, \ \forall i \in S$.\\

We now reduce the GAP to a special case of problem $P_B(S)$. We map the set of all agents (respectively, jobs) to the set of relays (respectively, files). We map the cost $c_{i,m}$ to the energy  $\frac{X_m}{R_{s,i}}P_{Rx,i}(R_{s,i})+\frac{X_m}{R_{Tx,i,S}}P_{Tx,i,S}(R_{Tx,i,S})$ spent by relay $i$ when it multicasts file $m$. Also, we map the profit $p_{i,m}$ to $\sum_{i \in S} d_{i,m}(-C_{r,i}^m(S))-\frac{aX_m}{R_{s,i}}P_{Rx,i}(R_{s,i})-\frac{aX_m}{R_{Tx,i,S}}P_{Tx,i,S}(R_{Tx,i,S}) + e$, which is the total energy costs incurred at relay $i$ and at the destination nodes in $S$ that request file $m$ when relay $i$ multicasts file $m$, plus a constant~\footnote{Note that the sum of the $U_{i,m}$ terms in \eqref{EQ:PS} is a  constant and hence these terms can be ignored. Also, the constant $e$ is chosen to be a large enough value so that all profits $p_{i,m}$ are mapped to non-negative values. Since $\sum_{i,m} e \alpha_{i,m}$ equals $e|\mathcal{M}_S|$, which is a constant, a constant gets added to the objective function due to the added $e$ terms; hence, the problem remains unchanged.} $e > 0$. Finally, we map the budget $t_i$ of agent $i$ to the maximum amount of energy $E_i$ that may be spent by relay $i$ (see \eqref{EQ:relay:max:energy:constraint}). 

With the above mapping, it can be checked that a feasible solution of the GAP instance with objective function value $\geq T$, for a given target $T$, exists iff a feasible solution of problem $P_B(S)$ with objective function value $\geq T^{\prime}$ for some target $T^{\prime}$ exists. The result follows.  
\ignore{(TODO: Define $\mathcal{M}_S$ to be the set of files requested by the users in $S$. In this subsection, and wherever appropriate in the rest of the paper, replace $\mathcal{M}$ with $\mathcal{M}_S$.)  }
\end{IEEEproof}

\subsection{Heuristics}
\label{SSC:model:B:heuristics}
We now provide some heuristics to solve problem $P_B(S)$. 
\subsubsection{Greedy Algorithm}
This algorithm is based on finding, for each file-user pair $(i\in S, m\in\mathcal{M}_S)$, the total energy cost that is incurred at all the cellular users (relay and destination nodes) if file $m$ is assigned to user $i$ for relaying to its destination nodes; let $C_{i,m}$ denote this energy cost.  The greedy algorithm sorts $C_{i,m}$, $i \in S$ for each file $m$ in increasing order. Then, starting from the most popular file~\footnote{The popularities of different files can be estimated using the history of file requests by different users in previous time slots.}, in decreasing order of file popularities, the algorithm assigns each file to the first user from its list of sorted users whose energy constraint is still met after the assignment. 

\subsubsection{Greedy Global Algorithm}
This algorithm is similar to an algorithm proposed in~\cite{RF:Martello}. 
In this algorithm, we calculate $C_{i,m}$ for all file-user pairs as in the greedy algorithm. For each file $m$, we construct a vector $\left(C_{m1,m},C_{m2,m},\ldots,C_{m|S|,m}\right)$ where $m1,\ldots,m|S| \in S$ and $C_{m1,m}\leq C_{m2,m}\leq\ldots\leq C_{m|S|,m}$. For each file $m$, we find the difference between $C_{m2,m}$ and $C_{m1,m}$ and select the file with the highest difference in the costs. Suppose file $\hat{m}$ has the highest cost difference. We assign file $\hat{m}$ to  user $\hat{m}1$ and remove file $\hat{m}$ from the list of files if the energy constraint of user $\hat{m}1$ is still met after the assignment. Otherwise, we remove the first element from the cost vector of file $\hat{m}$. Then we again find the file with the highest difference in costs between the second and first elements in its cost vector and repeat this process until a relay is assigned to each file.

\section{Numerical Results}
\label{SC:numerical:results}
We present numerical results in this section. 
In Section~\ref{SSC:numerical:results:coalition:formation}, we consider Model $A$ and partition the set of all cellular users into four different clusters as shown in Fig.~\ref{figure:clusters}. The numbers of users in the four clusters are equal. We compare the total energy consumption of the cellular users under the case where the cellular users of each cluster form a coalition among themselves with that where there are no coalitions and every user downloads the file it requests directly from the BS. In Section~\ref{SSC:numerical:results:model:b:heuristics}, we consider problem $P_B(S)$ with $S = \mathcal{N}$, which was shown to be NP-Complete in Section~\ref{SSC:model:B:NP:completeness}. Using numerical computations, we evaluate the performances of the greedy and greedy global heuristics, which were described in Section~\ref{SSC:model:B:heuristics}, and that of an algorithm in which requested files are randomly assigned to relays. 

Throughout this section, we consider a set of users located in a hexagonal cell of radius $300$ meters, with the BS  at the center of the cell. We consider
that the probabilities with which different files are requested by users follow the \emph{Zipf's distribution}; note that the Zipf's distribution has been found to closely approximate the measured file popularity frequencies in several studies, \emph{e.g.},~\cite{RF:Breslau:Zipf}. Under the Zipf's distribution, if  $\mathcal{M}=\{1,\ldots,M\}$ is the set of all files that may potentially be requested, the probability with which file $i$ is requested by  a user is given by $p_i=\frac{\left(\frac{1}{i}\right)^{r_c}}{\sum\limits_{k\in\mathcal{M}}\left(\frac{1}{k}\right)^{r_c}}$, 
where $r_c$ is called the Zipf exponent. (Note that the set of files that are \emph{actually} requested by users is a subset of $\mathcal{M}$.) For modelling the channel, we consider distance dependent path loss along with lognormal shadow fading. We also assume that the channel  adds additive white gaussian noise (AWGN) and undergoes Rayleigh fading. Table~\ref{table:parameters} shows the values of various parameters used in the numerical computations.   

\begin{table}
\centering
\caption{Parameters Used in the Numerical Computations}
\begin{tabular}{|c|c|}
\hline
Parameter & Value\\
\hline
Propagation Model & \begin{tabular}{@{}c@{}}
Path loss with lognormal \\ shadow fading and \\Rayleigh fading\end{tabular}\\
Noise power & -174dBm/Hz\\
Standard deviation for shadow fading & 8\\
Path loss Exponent & 3.3 \\
Transmission power of BS &  40 dBm  \\
Relay transmission power  & 350 mW \\
Relay receiving power & 250 mW  \\
User receiving power on D2D link & 200 mW \\
Bandwidth of relay node & 10 MHz\\
Bandwidth of destination node & 10 MHz\\
File Size & \begin{tabular}{@{}c@{}}Uniformly distributed 
\\in the range \{1,\ldots,10\} Mb\end{tabular}\\
 \hline
\end{tabular}
\label{table:parameters}
\end{table}

\subsection{Comparison of the Cooperation and No Cooperation Cases under Model A}
\label{SSC:numerical:results:coalition:formation}
\begin{center}
\begin{figure}[!hbt]
\centering
\includegraphics[scale=0.5]{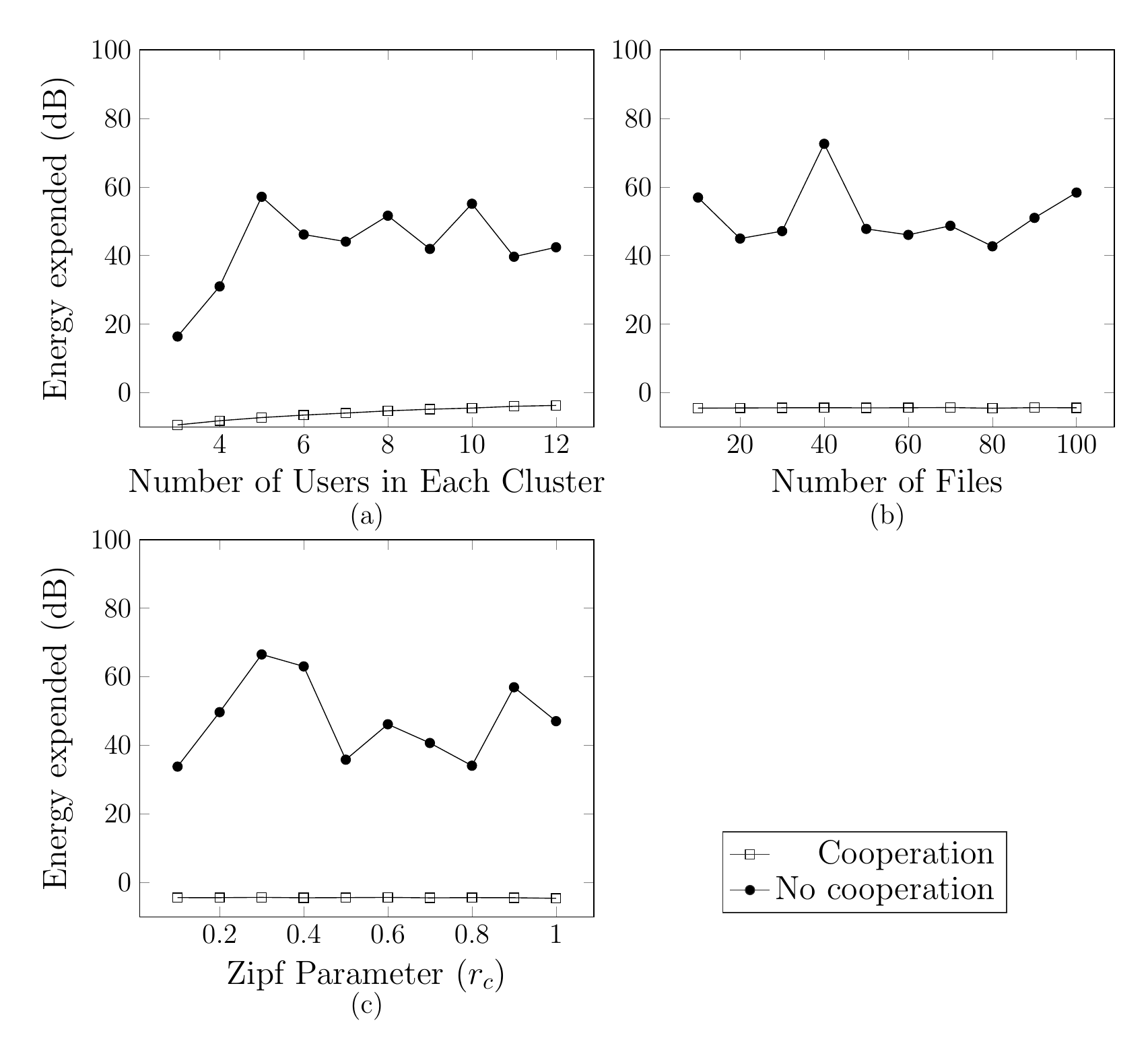}
\caption{\footnotesize{The figure shows various plots under the scenario where users are located in four clusters in a cell. Plot (a) shows the total energy expended versus the number of users in each cluster for the parameter values $M  = 50$ and  $r_c = 0.5$.  Plot (b)  shows the total energy expended versus the number of files, $M$, for the parameter values $N = 40$ and $r_c = 0.5$. Plot (c) shows the total energy expended versus the Zipf parameter, $r_c$, for the parameter values $M = 50$  and $N = 40$.}}
\vspace{-1em}
\label{FG:figure:cluster}
\end{figure}
\end{center}

For Model A, we analyse the case where the set of all cellular users are located in four clusters as shown in Fig.~\ref{figure:clusters}. The numbers of users in the four clusters are equal. Each cluster has a radius of 60 meters and the center of each cluster is located at a distance of 200 meters from the BS. The total energies expended by all the cellular users of the network in the case when cellular users of each cluster cooperate among themselves to form a coalition and in the case where each user directly downloads the file it requires from the BS are plotted versus the number of files, $M$,  number of users, $N$,  and the Zipf exponent, $r_c$, in Fig~\ref{FG:figure:cluster}.\footnote{In each plot, each point is obtained by taking an average over 100 runs.}  In all of the plots in Fig.~\ref{FG:figure:cluster}, the \emph{energy expended by users when they cooperate is much less than that when they act independently, which shows the benefits of cooperation}. 
\ignore{
Under scenario 1,  the gap between the expended energies in the ``cooperation'' and ``no cooperation'' cases increases with an increase in the number of users (see plot (a) of Fig.~\ref{FG:figure:random}). This is because, when the number of users is small, very few coalitions are formed since users are randomly distributed and hence the number of instances where  users are  located close to each other is low. The number of coalitions increases with an increase in the number of cellular users since the density of users in the cell increases.   On the other hand, in scenario 2, this gap is more pronounced even when the  number of users is small (see plot (a) in Fig.~\ref{FG:figure:cluster}). This is because coalition formation is possible in scenario 2) even for a small number of users since they are located in clusters. 
\begin{center}
\begin{figure}[!hbt]
\centering
\includegraphics[scale=0.5]{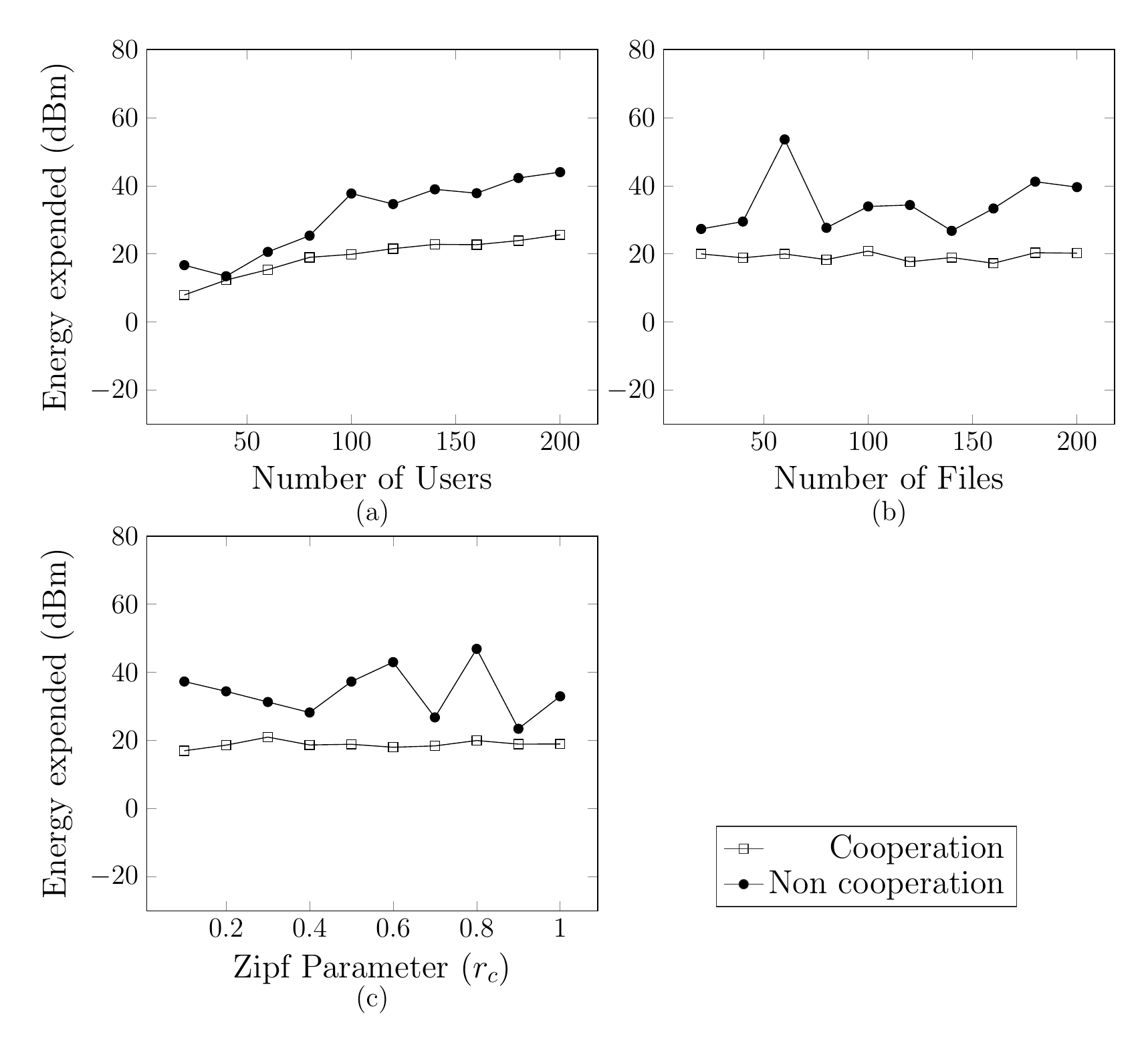}
\caption{\footnotesize{The figure shows various plots under the scenario where users are placed randomly in a cell. Plot (a) shows the total energy expended versus the number of users in each cluster for the parameter values $M  = 50$ and  $r_c = 0.5$.  Plot (b)  shows the total energy expended versus the number of files, $M$, for the parameter values $N = 40$ and $r_c = 0.5$. Plot (c) shows the total energy expended versus the Zipf parameter, $r_c$, for the parameter values $M = 50$  and $N = 40$.}}
\vspace{-1em}
\label{FG:figure:random}
\end{figure}
\end{center}}

\subsection{Performance of Heuristics under Model B}
\label{SSC:numerical:results:model:b:heuristics}
For Model B, we analyse the case where the cellular users are located randomly across the cell. The total energies expended by all the cellular users of the network (relays and destination nodes) under the greedy, greedy global and random algorithms were computed for various parameter values.

\begin{center}
\begin{figure}[!hbt]
\centering
\includegraphics[scale=0.5]{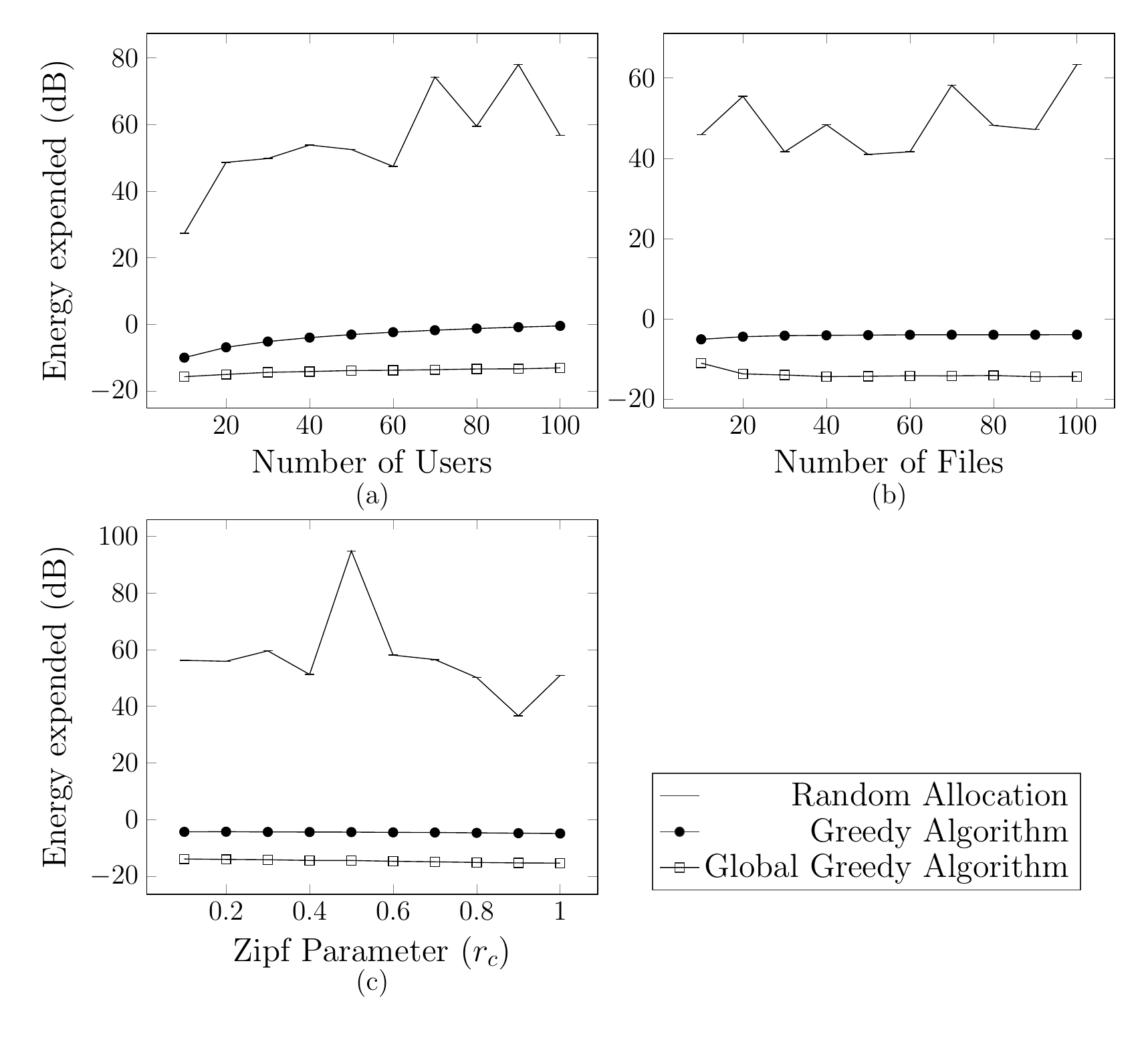}
\caption{\footnotesize{The figure shows various plots under the scenario where users are placed randomly in a cell. Plot (a) shows the total energy expended versus the number of users, $N$, for the parameter values $M  = 50$ and  $r_c = 0.5$.  Plot (b)  shows the total energy expended versus the number of files, $M$, for the parameter values $N = 40$ and $r_c = 0.5$. Plot (c) shows the total energy expended versus the Zipf parameter, $r_c$, for the parameter values $M = 50$  and $N = 40$.}}
\vspace{-1em}
\label{FG:figure}
\end{figure}
\end{center}

The total energy expended is plotted versus the number of users, $N$,  number of files, $M$,  and the Zipf exponent, $r_c$, in Fig.~\ref{FG:figure}. All three plots show that the energy expended under the greedy global algorithm is lower when compared to that under the greedy algorithm, which in turn is lower than that under the random algorithm. Intuitively, the greedy global algorithm performs better than the greedy algorithm since the former algorithm takes into account not only the costs $C_{k,i}$, but also the differences, $C_{i2,i} - C_{i1,i}$, during the allocation process. 
\vspace{-1em}
\ignore{
From  Fig.~\ref{FG:figure}(a), we see that, overall, the total energy expended increases in the number of users for all the three algorithms. Intuitively, this is because more users request files and also a file is multicast by a relay at the minimum of the achievable data rates between itself and all the destination nodes requesting the file; so as the number of randomly located destination nodes increases, data is sent at lower rates, which consumes more energy.  Also, from Fig.~\ref{FG:figure}(b), it can be seen that the total energy expended remains same despite an increase in the number of files for the greedy and global greedy algorithms.  Also, the curves corresponding to greedy and global greedy algorithm saturate when the number of files becomes large; this is because the newly added files are requested with very low probabilities under the Zipf distribution. Next, from  Fig.~\ref{FG:figure}(c), it can be seen that the total energy expended decreases in the Zipf parameter $r_c$ for all the three algorithms. Intuitively, this is because as $r_c$ increases, the distribution becomes more concentrated over a few popular files; hence, only a few (popular) files are requested with high probabilities and the remaining files are requested with low probabilities. }

\section{Conclusions}
\label{SC:conclusions:coalition}
We considered a scenario in which cellular users can employ relaying and use D2D
communication to transfer files requested by users from the BS and studied conditions under which users have an incentive to cooperate with each other. 
We considered two different relaying models: Model A
and Model B. First, we showed that, in general, the above coalitional game under Model A may have an empty core, i.e., it may not be possible to stabilize the grand coalition.  Next, we provided conditions under which 1) the core is always non-empty and 2) a $\mathbb{D}_c$-stable partition always exists. Also, we showed that under Model B, the problem of assigning relays to destination nodes so as to maximize the sum of utilities of all the users is NP-Complete. Our numerical results show that when cellular users cooperate with each other, the total amount of energy consumed in transferring the requested files from the BS to the destination nodes can be considerably reduced compared to the case when each user separately downloads the file it needs from the BS. 

\bibliographystyle{IEEEtran}  
\bibliography{sample} 

\end{document}